\documentclass[10pt, twocolumn]{IEEEtran}

\usepackage{multirow}
\usepackage{booktabs}
\usepackage{longtable}
\usepackage[justification=centering]{caption}
\usepackage{cite}

\usepackage{graphicx}
\usepackage{amsmath}
\usepackage{amsfonts}
\usepackage{amssymb}
\usepackage{stfloats}
\usepackage{arydshln}

\usepackage{subfigure}
\graphicspath{{figure/}}

\usepackage{amsfonts, amsmath, mathrsfs, amsbsy, amssymb, dsfont, color}
\usepackage{graphicx}

\newcommand\prefixtext[1]{%
  \ifvmode\else\\\@empty\fi
  \noalign{%
    \penalty0%
    \vbox{\mathstrut}%
    \penalty10000%
    \vskip-\baselineskip
    \penalty10000%
    \vbox to 0pt{%
      \normalbaselines
      \ifdim\linewidth=\columnwidth
      \else
        \parshape\@ne
        \@totalleftmargin\linewidth
      \fi
      \vss
      \noindent#1\par}%
      \penalty10000%
      \vskip-\baselineskip}%
      \penalty10000}

\newtheorem{lemma}{Lemma}

\newcommand{\qed}{\nobreak \ifvmode \relax \else
      \ifdim\lastskip<1.5em \hskip-\lastskip
      \hskip1.5em plus0em minus0.5em \fi \nobreak
      \vrule height0.75em width0.5em depth0.25em\fi}

\DeclareMathAlphabet{\mathpzc}{OT1}{pzc}{m}{it}
\newcommand{\comment}[1]{}
\def\({\left(}
\def\){\left)}

\def\bth{\boldsymbol{\theta}}

\def\bth{\boldsymbol{\theta}}

\def\bSi{\boldsymbol{\Sigma}}

\def\({\left(}
\def\){\right)}
\def\[{\left[}
\def\]{\right]}
\def\BEq{\begin{eqnarray}}
\def\EEq{\end{eqnarray}}
\def\BE*{\begin{eqnarray*}}
\def\EE*{\end{eqnarray*}}
\def\BA{\begin{array}}
\def\EA{\end{array}}

\def\Nn{\nonumber}
\def\0{\mathbf{0}}
\def\1{\mathbf{1}}

\def\A{\mathbf{A}}

\def\C{\mathbf{C}}

\def\D{\mathbf{D}}

\def\F{\mathbf{F}}

\def\I{\mathbf{I}}

\def\k{\mathbf{k}}

\def\l{\mathbf{l}}

\def\Q{\mathbf{Q}}

\def\R{\mathbf{R}}

\def\v{\mathbf{v}}

\def\w{\mathbf{w}}

\def\x{\mathbf{x}}
\def\X{\mathbf{X}}
\def\y{\mathbf{y}}

\def\z{\mathbf{z}}

\def\and{\prefixtext{and}}

\def\Im{\mathrm{Im}}
\def\Re{\mathrm{Re}}

\title{One-Bit Covariance Reconstruction with Non-zero Thresholds: Algorithm and Performance Analysis}
\author{Yu-Hang Xiao,~\IEEEmembership{Member,~IEEE}, Lei Huang,~\IEEEmembership{Senior Member,~IEEE}, David Ram{\'\i}rez,~\IEEEmembership{Senior Member,~IEEE}, Cheng Qian,~\IEEEmembership{Member,~IEEE}, and~Hing Cheung So,~\IEEEmembership{Fellow,~IEEE}
\thanks{\textcolor{blue}{This work has been submitted to the IEEE for possible publication.
Copyright may be transferred without notice, after which this version may
no longer be accessible.}}
\thanks{Y.-H Xiao and L. Huang are with the State Key Laboratory of Radio Frequency Heterogeneous Integration, Shenzhen University, Shenzhen 518060, China (e-mail: yuhangxiao@szu.edu.cn; dr.lei.huang@ieee.org).}
\thanks{David Ram{\'\i}rez is with the Department of Signal Theory and Communications, Universidad Carlos III de Madrid, Madrid 28903, Spain, and also with the Gregorio Mara{\~n}{\'o}n Health Research Institute, Madrid 28007, Spain (email: david.ramirez@uc3m.es).}
\thanks{C. Qian is with the IQVIA Inc., Cambridge, MA 02139, USA (e-mail: alextoqc@gmail.com).}
\thanks{H. C. So is with Department of Electronic Engineering, City University of Hong Kong, Hong Kong, China (e-mail: hcso@ee.cityu.edu.hk).}
\thanks{The work of Yu-Hang Xiao was supported by the National Natural Science Foundation of China under Grant 62201359. The work of Lei Huang was supported in part by the National Science Fund for Distinguished Young Scholars under Grant 61925108, and in part by the National Natural Science Foundation of China under Grants U1913221. The work of D. Ram{\'\i}rez was partially supported by MCIN/AEI/10.13039/501100011033/ FEDER, UE, under grant PID2021-123182OB-I00 (EPiCENTER) and by the Office of Naval Research (ONR) Global under contract  N62909-23-1-2002.}
}

\begin{document}

\input{epsf}
\date{}

\maketitle

\begin{abstract}


Covariance matrix reconstruction is a topic of great significance in the field of one-bit signal processing and has numerous practical applications. Despite its importance, the conventional arcsine law with zero threshold is incapable of recovering the diagonal elements of the covariance matrix. To address this limitation, recent studies have proposed the use of non-zero clipping thresholds. However, the relationship between the estimation error and the sampling threshold is not yet known. In this paper, we undertake an analysis of the mean squared error by computing the Fisher information matrix for a given threshold. Our results reveal that the optimal threshold can vary considerably, depending on the variances and correlation coefficients. As a result, it is inappropriate to use a constant threshold to encompass parameters that vary widely. To mitigate this issue, we present a recovery scheme that incorporates time-varying thresholds. Our approach differs from existing methods in that it utilizes the exact values of the threshold, rather than its statistical properties, to enhance the estimation performance. Our simulations, including the direction-of-arrival estimation problem, demonstrate the efficacy of the developed scheme, especially in complex scenarios where the covariance elements are widely separated.
\end{abstract}


\begin{sloppypar}
\section{Introduction}

One-bit analog-to-digital converters (ADCs) have garnered significant attention in recent years due to their unique merits over high-resolution ADCs. These advantages include cost-effectiveness, lower power consumption, and simpler hardware design. In addition, the reduced data flow associated with one-bit ADCs makes data storage and transmission more manageable. This has led to the widespread application of one-bit signal processing in various fields, such as multiple-input multiple-output communications~\cite{Balevi2019,Zhang2020,Mo2015,Qian2019SPL,Li2017,Choi2016TC}, array processing~\cite{Shalom2002TAES,Yu2016SPL,Liu2017ICASSP,Stein2016ITGWSA,Sedighi2020,Sedighi2021}, and radar~\cite{Ren2017,Xiao2022,Xi2020TAES,Xi2020TSP,Liu2021IET,Jin2020TAES,Stein2015TSP}.

Despite its numerous advantages, one-bit analog-to-digital conversion has created challenges in some common applications, such as parameter estimation and detection. The loss of amplitude information has limited its use in areas that rely on second-order statistics, such as direction-of-arrival (DOA) estimation~\cite{Qian2016}, spectrum sensing~\cite{Wei2012a,Xiao2018a,Zhao2021a}, and radar target detection~\cite{Xiao2018b,Liu2015TAES}. Therefore, the reconstruction of the covariance matrix has become a critical topic in one-bit processing research.


The most frequently employed criterion for recovering one-bit covariance matrices is the arcsine law~\cite{Vleck1966}, which is also referred to as an extension of the Bussgang theorem~\cite{Bussgang1952,Minkoff}. It may immediately translate the one-bit covariance matrix into that of the unquantized data matrix. It does, however, provide a normalized version of the covariance matrix, namely the correlation matrix,\footnote{This matrix contains all pairwise correlation coefficients.} rather than the original covariance matrix.
That is, unless the diagonal elements of the covariance matrix are equal, the estimation is biased and inconsistent.
The explanation for this phenomenon is that these systems use zero as the sampling threshold, meaning that the likelihood of the quantized signal has no bearing on the variance of the random variables. As a result, these samples cannot be used to estimate variances, i.e., the diagonal entries of the covariance matrix.

To address this issue,  Liu and Lin~\cite{Liu2021} have employed a constant (non-zero) threshold to enable accurate and consistent estimates of the covariance matrix, which may be easily accomplished by adding a DC level to the input signal.
With the addition of the non-zero threshold, the likelihood of the output being $+1$ or $-1$ is no longer fixed at $1/2$ but is instead  a function of the ratio between the threshold and the standard deviation of the random variable. This allows the variance to be estimated. Its extension to time-varying thresholds  is suggested in~\cite{Eamaz2022} by adding a random dithering signal to the constant threshold. This is equivalent to modifying the population covariance matrix of the signal prior to quantizing with a constant threshold.

However, there is still no performance analysis conducted to derive the estimation error associated with the threshold and the population covariance matrix, making it impossible to optimize the threshold value to improve estimation performance.
In addition, without such analysis, we cannot set the dithering signal properly to relocate the covariance matrix to an appropriate region.

In this paper, we analyze the performance of the constant threshold estimator in~\cite{Liu2021}, which is also compatible with the random threshold method in~\cite{Eamaz2022}. Due to the absence of closed-form estimators, it is prohibitive to define their statistical behavior using conventional methodologies. Our idea is to perform a Taylor's expansion and then use the result to compute the mean squared error (MSE) of the estimators. The result indicates that a low threshold facilitates the estimation of the non-diagonal elements while diagonal ones favour thresholds comparable to their square roots. Therefore, it is inappropriate to adopt a constant threshold to deal with all elements in the covariance matrix, especially when the parameters are distinct from each other, as is typical when the dimension increases.

To address this issue, we present a novel approach based on a time-varying threshold, which differs from~\cite{Eamaz2022} since it uses the exact values of the threshold and not only its statistical properties. Using Price's theorem~\cite{Price1958}, we calculate the gradient of the orthant probability with regard to the covariance matrix parameters and seek the maximum likelihood estimators (MLEs) of the parameters. The algorithm is also extended to complex-valued scenarios to accommodate array processing applications. Furthermore, we carry out performance analysis of the new method by computing the inverse of the Fisher information matrix, which allows us to predict the performance more efficiently than through Monte Carlo simulations.

Finally, simulation results are presented to demonstrate the effectiveness of our proposed approach.
We consider the direction-of-arrival (DOA) estimation of coherent sources, which requires the reconstruction of the received signals covariance matrix, as an example.  We first estimate the covariance matrix through different methods and then process the results with the Enhanced Principal-singular-vector Utilization for Modal Analysis (EPUMA)~\cite{Qian2016} algorithm to produce DOA estimates.
It is shown that compared to constant and random threshold-based methods, our algorithm achieves significantly improved accuracy and stability.

The key contributions of this paper are as follows:

\begin{enumerate}

\item We conduct a thorough performance analysis of the constant threshold approach by leveraging a Taylor's expansion to analyze the estimator, indicating that it is challenging to use a constant threshold to effectively estimate parameters distributed over a wide range. This finding opens up the opportunity for optimization of the sampling threshold.

\item We introduce a new sampling strategy that utilizes time-varying thresholds and the corresponding recovery algorithm. In comparison to the existing constant and random threshold approaches, our solution offers higher estimation accuracy and demonstrates improved robustness against parameter unevenness  and high correlation coefficients.

\item To further analyze the algorithm performance, we compute the Fisher information corresponding to each threshold value. Our results demonstrate that the Fisher information provides a precise performance indicator even when the likelihood function is inconsistent across different samples.

\item Finally, we extend the covariance matrix estimator to the complex-valued scenario and integrate it with the EPUMA for DOA estimation, highlighting the broad range of potential applications.

\end{enumerate}

In Section II we formulate the problem and review the related works for one-bit covariance estimation, and in Section III, we analyze their performance. Section IV presents our novel estimator and investigates its statistical behavior. Section V demonstrates the usefulness and effectiveness of our estimator by combining it with EPUMA for DOA estimation, where we also conduct simulations to corroborate our theoretical calculations.

\subsection*{Notation}

Throughout this paper, we use boldface uppercase letters for matrices, boldface lowercase letters for column vectors, and
lightface lowercase letters for scalar quantities. The notation $\A\in\mathbb{R}^{p\times q} \ (\mathbb{C}^{p\times q})$ indicates that $\A$ is a $p\times q$ real (complex) matrix.
The operators $\mathbb{E}[a]$ and $\mathbb{V}[a]$ denote, respectively, the expectation and variance of random variable $a$, $\mathbb{C}[a,b]$ is the covariance between $a$ and $b$, and $\sim$ means ``distributed as''.
The superscript $\hat{a}$ denotes the estimate of $a$.
Finally, the operators $\operatorname{Re}(\cdot)$ and $\operatorname{Im}(\cdot)$ extract the real and imaginary parts of their argument and $\imath=\sqrt{-1}$ is the imaginary unit.

\section{Preliminaries}
\label{sec:problem formulation}
In this section, we present the problem of one-bit covariance estimation and review existing methods based on various sampling schemes, including the zero threshold, constant threshold, and random threshold approaches.

\subsection{Problem Formulation}

Suppose $\y\in\mathbb{R}^{M\times 1}$ follows a zero-mean multivariate Gaussian distribution $\mathcal{N}(\0,\bSi_{\y})$. Assume we have $N$ i.i.d. one-bit quantized observations of $\y$:
\begin{align}
\x(t)=\mathrm{sign}(\y(t)-\v(t)),~~~t=1,\cdots,N,
\end{align}
where
\begin{align}
\x(t)&=[x_1(t),\cdots,x_M(t)]^T, \\
\y(t)&=[y_1(t),\cdots,y_M(t)]^T,
\end{align}
and
\begin{align}
\v(t)=[v_1(t),\cdots,v_M(t)]^T,
\end{align}
is the quantization threshold vector. The function $\mathrm{sign}(\cdot)$ is the quantization operator
\begin{align}
\mathrm{sign}(x)=
\begin{cases}
+1, & x\geq 0,\\
-1, & x<0. \\
\end{cases}
\end{align}
Our aim is to recover the covariance matrix of the unquantized signal $\y$:
\begin{align}
\bSi_{\y}=\mathbb{E}[\y\y^T],
\end{align}
given its one-bit quantized sample, i.e., $\X=[\x(1),\cdots,\x(N)]$.
To simplify our discussion, we focus on the $2\times 2$ case:
\begin{align}
\bSi_{\y}=
\begin{bmatrix}
\sigma_1^2~~  \sigma_{12}\\
\sigma_{12}~~ \sigma_2^2
\end{bmatrix},
\end{align}
which can be easily extended to the general case in a pairwise manner.

There are various methods of setting the threshold $\v(t)$.
Traditionally, it is fixed at $\v(t) = \0$, resulting in the complete loss of amplitude information while only the correlation coefficients can be obtained. In order to estimate the variance of the random variables, it is necessary to set $\v(t)$ to be non-zero by incorporating a control sequence at the input of the ADC. This control sequence can be a DC level~\cite{Liu2021}, or taking a time-varying form, such as a sine wave~\cite{Zhao2019} or a randomly generated sequence~\cite{Eamaz2022,Knudson2016}.

\subsection{Zero Threshold}

When the sampling threshold is $0$, the relationship between $\bSi_{\x}$ and $\bSi_\y$ can be described using the well-known \emph{arcsine law}~\cite{Vleck1966}:
\begin{align}
\bSi_{\x}=\frac{2}{\pi}\sin^{-1} \left(\D_{\y}^{-\frac{1}{2}}\bSi_{\y}\D_{\y}^{-\frac{1}{2}}\right),
\end{align}
where $\D_{\y}=\text{diag}(\bSi_{\y})$. Assuming that $\D_{\y}$ is the identity matrix, a natural estimator of $\bSi_{\y}$ is
\begin{align}
\hat{\bSi}_{\mathbf{y}}=\sin\(\frac{\pi}{2}\hat{\bSi}_{\mathbf{x}}\),
\end{align}
where $\hat{\bSi}_{\x}$ is the sample covariance matrix of $\x$:
\begin{align}
\hat{\bSi}(\x)=\frac{1}{N}\sum_{t=1}^N\x(t)\x(t)^T.
\end{align}
In the complex-valued case, where the sampling process is modified as
\begin{align}\label{Y}
\x=\mathcal{Q}(\y) = \mathrm{sign}(\Re(\y)-\v) + \imath \mathrm{sign}(\Im(\y)-\v),
\end{align}
the estimator is modified accordingly as
\begin{align}\label{complex_arcsine_law}
\hat{\bSi}_{\mathbf{y}}=\sin\(\frac{\pi}{4}\Re(\hat{\bSi}_{\mathbf{x}})\)+\imath \sin\(\frac{\pi}{4}\Im(\hat{\bSi}_{\mathbf{x}})\).
\end{align}
Interestingly, the work~\cite{Liu2020} demonstrated that \eqref{complex_arcsine_law} holds not only for  complex circular Gaussian distributions, but all complex elliptically symmetric distributions. However, a significant drawback of the arcsine law is that it is incapable of estimating the diagonal entries of $\bSi_{\mathbf{y}}$, as the likelihood function does not include these entries. That said, if the assumption of unit diagonal entries is violated, the arcsine law becomes biased and inconsistent.

\subsection{Constant Threshold Approach}

The use of constant threshold has been introduced in~\cite{Liu2021} for covariance matrix recovery.
The reconstruction can be accomplished based on the following probabilities:
\begin{align}
    p_{i}&=\Pr\{x_i = +1\}=Q\(\frac{v}{\sigma_i}\),\,\,\,\,\,i=1,2,\\
    p_{12}&=\Pr\{x_1 = +1,x_2 = +1\}\Nn\\
          &=\int_{\frac{v}{\sigma_1}}^\infty\int_{\frac{v}{\sigma_2}}^\infty f\left(y_1,y_2\Big|\frac{\sigma_{12}}{\sigma_1\sigma_2}\right)dy_1 dy_2\label{p_hat},
\end{align}
where $v$ is the threshold, $f(y_1,y_2|\rho)$ is the probability density function of bivariate Gaussian distribution with unit variances and correlation coefficient $\rho$, given by
\begin{align}
f(y_1,y_2|\rho)=\frac{1}{2\pi\sqrt{1-\rho^2}}\exp\left(-\frac{ y_1^2-2\rho y_1 y_2+ y_2^2}{2(1-\rho^2)}\right),
\end{align}
and
\begin{align}
Q(a)=\int_a^{\infty}\frac{1}{\sqrt{2\pi}}\exp\(\frac{-t^2}{2}\)dt.
\end{align}
The MLEs of the probabilities are:
\begin{align}\label{p_MLE}
\hat{p}_i&=\frac{\sum_{t=1}^N [x_i(t)+1]}{2N},\,\,\,\,\,i=1,2,\\
\hat{p}_{12}&=\frac{\sum_{t=1}^N [x_1(t)+1][x_2(t)+1]}{4N}.
\end{align}
As a consequence, and using the invariance property of the MLE, the MLEs of the variances are
\begin{align}
\label{eq:estimator_variance}
\hat{\sigma}_i=\frac{v}{Q^{-1}(\hat{p}_i)},\,\,\,\,\,i=1,2.
\end{align}
On the other hand, the right hand side of \eqref{p_hat} can be rewritten as the following infinite polynomial form:\footnote{Note that the result here is slightly modified, as opposed to the original version in~\cite{Liu2021}, to cope with the non-uniform variances.}
\begin{align}
\hat{p}_{12}  = \frac{e^{-\frac{v^2}{\hat{\sigma}_{1}\hat{\sigma}_{2}}}}{\pi}
\sum_{k =0}^{\infty} \frac{H_{k}\left(\frac{v}{\sqrt{2} \hat{\sigma}_{1}}\right)H_{k}\left(\frac{v}{\sqrt{2} \hat{\sigma}_{2}}\right)}{2^{k+1}(k+1)!} \rho^{k+1}+\frac{\mu_{1}\mu_{2}}{4},
\end{align}
where
\begin{align}
\mu_{i}&=2 Q\left(\frac{v}{\hat{\sigma}_{i}}\right)-1,\,\,\,\,\,i=1,2,
\end{align}
and
\begin{align}
H_{k}(a)&=(-1)^{k} e^{a^{2}}\frac{d^k}{da^k} e^{-a^{2}},
\end{align}
is the Hermite polynomial of order $k$. The correlation coefficient $\rho$ can then be estimated numerically by solving the equation omitting higher-order terms of the polynomial.

\subsection{Random Threshold}
In~\cite{Eamaz2022}, the use of a random threshold with a Gaussian distribution $\mathcal{N}(v\1_M,\bSi_t)$ is suggested. This is equivalent to shifting the original covariance matrix to $\bSi'=\bSi+\bSi_t$ and estimating $\bSi'$ with a constant sampling threshold $v\1_M$.
Although~\cite{Eamaz2022} adopts a different numerical method to solve the MLE of the covariance as opposed to~\cite{Liu2021}, the difference does not affect the statistical efficiencies. 

In general, non-zero threshold approaches surpass the arcsine law as they allow for the full recovery of the covariance matrix. However, it remains unclear whether a constant threshold is optimal. Particularly, no performance analysis has been conducted to determine whether estimating $\bSi$ or $\bSi'$ provides smaller MSE, which makes it impossible to determine the shifting matrix $\bSi_t$.  In addition, it is unknown which threshold provides optimum estimation for different diagonal and non-diagonal elements.

In this paper, we first analyze the MSE of the constant threshold estimator, revealing that the optimal threshold for estimating different variances and covariances are distinct. We then present a recovery algorithm based on time-varying thresholds, where the thresholds are known deterministic values instead of random variables, as opposed to~\cite{Eamaz2022}. 


%

\section{Performance Analysis of Constant Threshold Approach}
\label{sec:performance}


In this section, we analyze the MSE of the constant-threshold-based approach with regard to both variance and covariance estimations. The analysis is conducted by applying a Taylor's expansion to the expressions of the estimators. For the estimation of variances, a Taylor's expansion up to the second order is applied, while for the estimation of covariances, a first-order expansion is employed due to the complexity of the estimator.


\subsection{MSE of Diagonal Elements}

The approximation is made under the assumption that $N$ is large, which is a common scenario in one-bit systems as the sampling rate is typically very high. Furthermore, as it has been proved that the bias of MLE approaches $0$ as $N\rightarrow \infty$~\cite{Kay_estimation}, the MSE of the detector becomes equivalent to the variance of the estimators.

Recall that the variance estimator for $\sigma_i$ $(i=1,2)$ is
\begin{align}
\hat{\sigma}_i=\frac{v}{Q^{-1}(\hat{p}_i)}.
\end{align}
We first compute the second-order Taylor's expansion of the estimator. For simplicity, we define
\begin{align}
h(a)=\frac{v}{Q^{-1}(a)}.
\end{align}
The second-order Taylor's expansion of $h(a)$ at $a=p_i$ is:
\begin{multline}\label{expansion_var}
h(a)=h(p_i)+h'(p_i)(a-p_i) \\ + \frac{1}{2}h''(p_i)(a-p_i)^2+\mathcal{O}((a-p_i)^3),
\end{multline}
where
\begin{align}
h'(p_i)
=&\frac{\sqrt{2\pi}\sigma_i^2}{v}\exp\(\frac{v^2}{2\sigma_i^2}\),\\
h''(p_i)=&\exp\(\frac{v^2}{\sigma_i^2}\) \left(\frac{4\pi\sigma_i^3}{v^2} - 2\pi \sigma_i \right).
\end{align}
\begin{IEEEproof}
    See Appendix \ref{appendix:A}.
\end{IEEEproof}

According to \eqref{expansion_var}, the variance of $\hat{\sigma}_i$ can be approximated as:
\begin{multline}\label{expansion_var_1}
\mathbb{V}(\hat{\sigma}_i) =
(h'(p_i) - h''(p_i) p_i)^2 \mathbb{V}(\hat{p}_i)  + \frac{1}{4}[h''(p_i)]^2 \mathbb{V} (\hat{p}_i^2) \\ + (h'(p_i) - h''(p_i) p_i) h''(p_i) \mathbb{C}(\hat{p}_i,\hat{p}_i^2).
\end{multline}
Next, we calculate the terms $\mathbb{V}(\hat{p}_i)$, $\mathbb{V}(\hat{p}_i^2)$ and $\mathbb{C}(\hat{p}_i,\hat{p}_i^2)$, which requires us to first compute the second- to fourth-order moments of $\hat{p}_i$.
Since $N_i = N \hat{p}_i$ follows a binomial distribution,  its moments can be evaluated by the following lemma~\cite{Knoblauch2008}.
\begin{lemma}\label{lemma:moments}
The $c$th order moment of a binomial distributed random variable $\vartheta$ with success probability $p_i$ and number of trials $N$ is:
\begin{align}
\mathbb{E}[\vartheta^c]=\sum_{k=0}^{c}S_k^c N^{\underline{k}} p_i^{k},
\end{align}
where $S_k^c$ is the  Stirling number of the second kind:
\begin{align}
S_k^c = \sum_{j=1}^k(-1)^{k-j}\frac{j^{c-1}}{(j-1)!(k-j)!}, 
\end{align}
and $N^{\underline{k}}$ is the $k$th falling power of $N$:
\begin{align}
N^{\underline{k}}=N(N-1)\cdots(N-k+1).
\end{align}
\end{lemma}
Using Lemma~\ref{lemma:moments} with $\vartheta = N \hat{p}_i$, the required moments of $\hat{p}_i$ are:
\begin{align}
m_2\!=\!\mathbb{E}[\hat{p}_i^2]&=\frac{p_i+p_i^2(N-1)}N,\\
m_3\!=\!\mathbb{E}[\hat{p}_i^3]&=\frac{p_i+3p_i^2(N-1)+p_i^3(N-1)(N-2)}{N^2},\\
m_4\!=\!\mathbb{E}[\hat{p}_i^4]&=\frac{p_i+7p_i^2(N-1)+6p_i^3(N-1)(N-2)}{N^3}\Nn\\
&\phantom{=} +\frac{p_i^4(N-1)(N-2)(N-3)}{N^3},
\end{align}
where $m_k$ denotes the $k$th order moment of $\hat{p}_i$.
Therefore, we have
\begin{align}
\mathbb{V}(\hat{p}_i)&=m_2-p_i^2, \label{variances}\\
\mathbb{V}(\hat{p}_i^2)&=m_4-m_2^2,\\
\mathbb{C}(\hat{p}_i,\hat{p}_i^2)&=m_3 - p_i m_2 \label{cross_variances}.
\end{align}
Substituting \eqref{variances}-\eqref{cross_variances} into \eqref{expansion_var_1} results in the variance of $\hat{\sigma}_i$.

\subsection{MSE of Non-Diagonal Elements}

The analysis of the covariance estimator is more complex compared to the variance estimator as it depends not only on $\hat{p}_{ij}$, but also on the estimated variances $\hat{\sigma}_i$ and $\hat{\sigma}_j$. Therefore, a second-order analysis is not feasible and a first-order analysis is conducted instead. This involves constructing a linear approximation of $\hat{\sigma}_{ij}$, resulting in a simplified representation of its behavior. The result is summarized in the following lemma.

\begin{lemma}\label{lemma:taylor_expansion}
The first-order Taylor's expansion of $\sigma_{12}$ as a function of $\hat{p}_1$, $\hat{p}_2$ and $\hat{p}_{12}$ is
\begin{align}
\sigma_{12}-\hat{\sigma}_{12}
=\l\[p_1-\hat{p}_1,p_2-\hat{p}_2,p_{12}-\hat{p}_{12}\]^T,
\end{align}
where
\begin{align}
\l=
\[-\frac{\partial \sigma_{12}}{\partial p_{12}}\frac{\partial p_{12}}{\partial \sigma_1}h'(p_1),-\frac{\partial \sigma_{12}}{\partial p_{12}}\frac{\partial p_{12}}{\partial \sigma_2}h'(p_2), \frac{\partial \sigma_{12}}{\partial p_{12}}\],
\end{align}
with
\begin{align}
\frac{\partial p_{12}}{\partial \sigma_{12}}
&= \left[\frac{\partial \sigma_{12}}{\partial p_{12}}\right]^{-1} =\frac{1}{\sigma_1\sigma_2}f\(\frac{v}{{\sigma}_1},\frac{v}{{\sigma}_2}\Big|\rho\), \\
\frac{\partial p_{12}}{\partial \sigma_1}
&=\frac{1}{\sigma_1}g\(\frac{v}{\sigma_1},\frac{v}{\sigma_2},\rho\)\label{derivative_sigma_1},\\
\frac{\partial p_{12}}{\partial \sigma_2}
&=\frac{1}{\sigma_2}g\(\frac{v}{\sigma_2},\frac{v}{\sigma_1},\rho\),\label{derivative_sigma_2}
\end{align}
where
\begin{align}\label{g_def}
    \!\!\!\!\!\!g&(\kappa_1,\kappa_2,\varrho)\Nn\\
    &=\frac{\kappa_1}{\sqrt{2\pi}}\exp\(-\frac{\kappa_1^2}{2}\)Q\left(\frac{\kappa_2-\varrho \kappa_1}{\sqrt{1-\varrho^2}}\!\right)-\varrho f\left(\kappa_1,\kappa_2|\varrho\right).
\end{align}
\end{lemma}
\begin{IEEEproof}
    See Appendix \ref{appendix:taylor_expansion}.
\end{IEEEproof}

Then, the variance of $\hat{\sigma}_{12}$ can be computed as
\begin{align}\label{cov_var}
\mathbb{V}[\hat{\sigma}_{12}]=\l\R\l^T,
\end{align}
where $\R$ is the covariance matrix of the random vector $[\hat{p}_1,\hat{p}_2,\hat{p}_{12}]^T$, which is
\begin{equation}\label{R}
\R=
\frac{1}{N}\begin{bmatrix}
p_1\bar{p}_1 & p_{12}-p_1p_2 & p_{12}\bar{p}_{1}  \\
p_{12}-p_1p_2 & p_2\bar{p}_2 & p_{12}\bar{p}_{2} \\
p_{12}\bar{p}_{1} & p_{12}\bar{p}_{2} & p_{12}\bar{p}_{12}
\end{bmatrix},
\end{equation}
where $\bar{p}_1=1-p_1$, $\bar{p}_2=1-p_2$ and $\bar{p}_{12}=1-p_{12}$.

\begin{IEEEproof}
    See Appendix \ref{appendix:C}.
\end{IEEEproof}

Substituting \eqref{R} into \eqref{cov_var} yields the variance of $\hat{\sigma}_{12}$.

\begin{figure}[!t]
  \includegraphics[width=0.9\columnwidth]{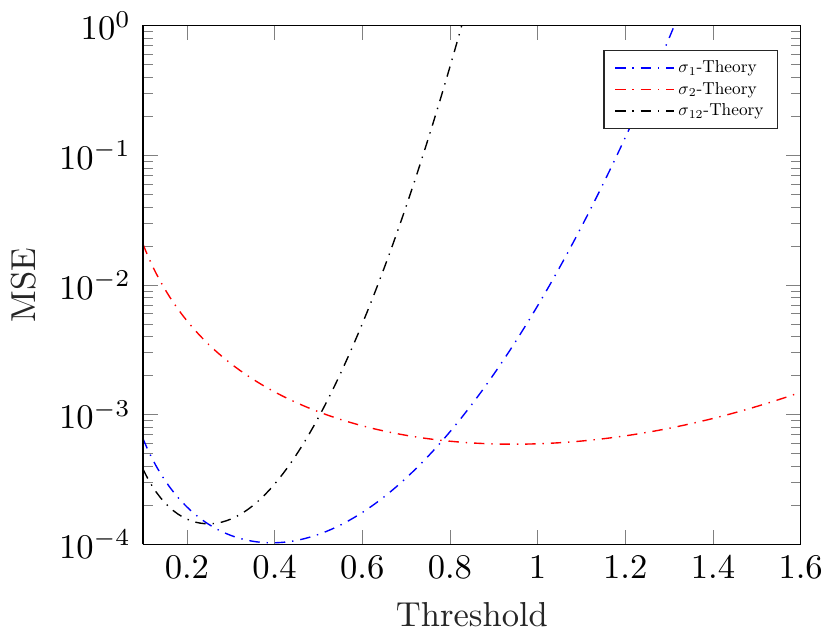}
\caption{Mean squared error versus threshold. }\label{fig1}
\end{figure}

Having obtained the theoretical performance of the constant-threshold estimator, we now conduct a simulation to study how the estimation error fluctuates with regard to the threshold value. In Fig. \ref{fig1}, we set $\sigma_1=0.25$, $\sigma_2=0.6$, $\sigma_{12}=-0.08$  and threshold varies from $0.1$ to $1.6$. It is clearly demonstrated that the optimal threshold for the three parameters can vary significantly. In this simulation, the optimal threshold value for the standard deviation estimation is approximately 1.6 times the population standard deviation, whereas the estimation of covariance prefers a low threshold. As a result, it is difficult to use a single threshold to deal with all the parameters. This issue is further compounded in real-world applications, where the parameters may be distributed over a broad range as the dimension increases. Consequently, recovery schemes incorporating time-varying thresholds are needed.

\section{Proposed Covariance Recovery Scheme}
\label{sec:detector}

In this section, we propose the implementation of a time-varying, known sampling threshold in lieu of constant or random sampling thresholds. Specifically, the sampling period is divided into $l$ sub-intervals of length $n$, with each sub-interval employing a distinct constant threshold. Compared to~\cite{Liu2021,Eamaz2022}, our approach has the potential to increase  robustness, particularly in situations where the diagonal entries differ significantly or the correlation coefficients are high.  To achieve this, we first establish the MLEs of $\sigma_1$ and $\sigma_2$ using the data from their respective channels, and then search for the MLE of $\sigma_{12}$ with the previously estimated $\sigma_1$ and $\sigma_2$ fixed. Then, the obtained values are used as the starting point for an iteration process that ultimately yields the joint MLE of $\bth=[\sigma_1,\sigma_2,\sigma_{12}]^T$. Finally, we prove that the joint MLE is numerically close to initial estimates when the number of sub-intervals is small. In such cases, we can omit using the joint MLE with negligible performance loss.

\subsection{Diagonal Entries}
\label{sec:rec_diagonal}

Without loss of generality, we study the MLE of $\sigma_i$ based on $\x_i=[x_i(1),\cdots,x_i(N)]$. The log-likelihood of $\sigma_i$ can be written as:
\begin{equation}\label{likelihood}
\mathcal{L}(\x_i;\sigma_i)=\sum_{t=1}^{N} \log\(Q\[\frac{x_i(t)v_i(t)}{\sigma_i}\]\).
\end{equation}
Consequently, the MLE of $\sigma_i$ is the solution of the following equation:
\begin{equation}
\frac{\partial{\mathcal{L}(\x_i;\sigma_1)}}{\partial{\sigma_i}}=\sum_{t=1}^{N}\frac{\Delta_{1,t}(\sigma_i)}{q_{i,t}(\sigma_i)}=0,
\end{equation}
where
\begin{align}
    \Delta_{1,t}(\sigma_i)&=\frac{v_i(t)}{\sqrt{2\pi}{\sigma}_i^2}\exp\(-\frac{v_i^2(t)}{2{\sigma}_i^2}\), \\
    q_{i,t}(\sigma_i)&=\frac{x_i(t)-1}{2}+p_{i,t}(\sigma_i),
\end{align}
with
\begin{equation}
    p_{i,t}(\sigma_i)=Q\(\frac{v_i(t)}{{\sigma}_i}\).
\end{equation}
We then obtain the ML estimate of $\sigma_i$ by the following Newton's iteration:
\begin{align}
\hat{\sigma}_{i}^{(l+1)}=\hat{\sigma}_{i}^{(l)}-\frac{\partial{\mathcal{L}(\x_i;\sigma_i)}}{\partial{\sigma_i}}\Big/\left.\frac{\partial^2{\mathcal{L}(\x_i;\sigma_i)}}{\partial\sigma_{i}^2}\right|_{\sigma_{i}=\hat{\sigma}_{i}^{(l)}},
\end{align}
where the second-order derivative is calculated as:
\begin{align}
\frac{\partial^2{\mathcal{L}(\x_i;\sigma_i)}}{\partial\sigma_{i}^2}=\sum_{t=1}^N\frac{q_{t}({\sigma}_i)\Delta_{2,t}({\sigma}_i)-\Delta_{1,t}^2({\sigma}_i)}{q_{t}^2({\sigma_i})},
\end{align}
with
\begin{align}
    \Delta_{2,t}(\sigma_i)=\frac{v_i^3(t)-2v_i(t)\sigma_i^2}{\sqrt{2\pi} \sigma_i^5}\exp\(-\frac{v_i^2(t)}{2\sigma_i^2}\).
\end{align}

\subsection{Non-Diagonal Entries}
\label{sec:rec_nondiagonal}

After obtaining the MLEs of $\sigma_1$ and $\sigma_2$, the covariance $\sigma_{12}$ can be estimated by assuming $\sigma_1=\hat{\sigma}_1$ and $\sigma_2=\hat{\sigma}_2$.
Therefore, we have
\begin{align}
p_{12,t}(\tilde{\rho})=\int_{\frac{v_1(t)}{\hat{\sigma}_1}}^\infty\int_{\frac{v_2(t)}{\hat{\sigma}_2}}^\infty f\left(y_1,y_2\Big|\tilde{\rho}\right)dy_1dy_2.\end{align}
where ${\tilde{\rho}}={{\sigma}_{12}}/(\hat{\sigma}_{1}\hat{\sigma}_{2})$.
According to the Price theorem~\cite{Price1958,Liu2021}, the derivative of $p_{12}$ with respect to $\tilde{\rho}$ is calculated as:
\begin{align}\label{derivative_rho}
    \frac{\partial p_{12,t}(\tilde{\rho})}{\partial \tilde{\rho}}=f\(\frac{v_1(t)}{\hat{\sigma}_1},\frac{v_2(t)}{\hat{\sigma}_2}\Big|\tilde{\rho}\).
\end{align}
Then, the log-likelihood function is
\begin{equation}
    \mathcal{L}(\X;\tilde{\bth}) = \sum_{t=1}^N\log\(o_t(\tilde{\bth})\),
\end{equation}
where $\tilde{\bth}=[\hat{\sigma}_1,\hat{\sigma}_2,\sigma_{12}]^T$ and
\begin{equation}
    o_t(\tilde{\bth})\!=\!\begin{cases}
p_{12,t}(\tilde{\rho}), & \!\!\! \x(t)=[+1,\!+1]^T\\
p_{1,t}(\hat{\sigma}_1)-p_{12,t}(\tilde{\rho}), & \!\!\! \x(t)=[+1,\!-1]^T\\
p_{2,t}(\hat{\sigma}_2)-p_{12,t}(\tilde{\rho}), & \!\!\! \x(t)=[-1,\!+1]^T\\
1\!-\!p_{1,t}(\hat{\sigma}_1)\!-\!p_{2,t}(\hat{\sigma}_2)\!+\!p_{12,t}(\tilde{\rho}), & \!\!\! \x(t)=[-1,\!-1]^T.\\
\end{cases}
\end{equation}
The first-order derivative of the log-likelihood is
\begin{equation}\label{de1_covariance}
    \frac{\partial{\mathcal{L}(\X;\tilde{\bth})}}{\partial{\sigma_{12}}}=\sum_{t=1}^N \frac{\Delta'_{1,t}({\rho})}{o_t(\tilde{\bth})}
\end{equation}
where
\begin{equation}
    \Delta'_{1,t}({\tilde{\rho}})=\frac{x_1(t)x_2(t)f\(w_1(t),w_2(t)\Big|{\tilde{\rho}}\)}{\hat{\sigma}_1\hat{\sigma}_2},
 \end{equation}
with
\begin{align}
w_1(t)&=\frac{v_1(t)}{\hat{\sigma}_1},&
w_2(t)&=\frac{v_2(t)}{\hat{\sigma}_2}.
\end{align}
In addition, the second-order derivative can be computed as
\begin{equation}
    \frac{\partial^2{\mathcal{L}(\X;\tilde{\bth})}}{\partial\sigma_{12}^2}=\sum_{t=1}^N\frac{o_t(\tilde{\bth})\Delta_{2,t}'({\tilde{\rho}})-[\Delta_{1,t}'({\tilde{\rho}})]^2}{o_t^2(\tilde{\bth})},
\end{equation}
where
\begin{align}
    \Delta_{2,t}'(\tilde{\rho})=&\frac{1}{2\pi\hat{\sigma}_1\hat{\sigma}_2\sqrt{1-{\tilde{\rho}}^2}}\left[\frac{{\tilde{\rho}}+w_1(t)w_2(t)}{1-{\tilde{\rho}}^2}-\frac{{\tilde{\rho}} u_t(\tilde{\rho})}{(1-{\tilde{\rho}}^2)^2}\right]\Nn\\
    &\times\exp\[-\frac{u_t(\tilde{\rho})}{2(1-{\tilde{\rho}}^2)}\],
\end{align}
with
\begin{equation}
    u_t(\tilde{\rho})=w_1^2(t)+w_2^2(t)-2{\tilde{\rho}} w_1(t)w_2(t).
\end{equation}
Similarly, we construct the Newton's iteration algorithm to solve this problem, which is:
\begin{equation}
\hat{\sigma}_{12}^{(l+1)}=\hat{\sigma}_{12}^{(l)}- \left[\frac{\partial^2{\mathcal{L}(\X;\tilde{\bth})}}{\partial\sigma_{12}}\Big/\left.\frac{\partial^2{\mathcal{L}(\X;\tilde{\bth})}}{\partial\sigma_{12}^2}\right]\right|_{\sigma_{12}=\hat{\sigma}_{12}^{(l)}}.
\end{equation}

\subsection{Joint MLE 
}
\label{sec:joint_rec}


Having obtained the initial estimates, we now seek the joint MLE of $\sigma_1$, $\sigma_2,$ and $\sigma_{12}$, which can be achieved using a gradient descent approach. Following the argument in \eqref{derivative_sigma_1} and \eqref{derivative_sigma_2}, it is easy to obtain the gradients of the log-likelihood with respect to $\sigma_1$ and $\sigma_2$ as
\begin{align}\label{derivative1}
\frac{\partial\mathcal{L} (\X;\bth)}{\partial \sigma_1}&= \sum_{t = 1}^N\frac{1}{\sigma_1o_t(\bth)}g\big({z_1(t)},{z_2(t)},x_1(t)x_2(t)\rho\big),\\
\label{derivative2}
\frac{\partial\mathcal{L} (\X;\bth)}{\partial \sigma_2}&= \sum_{t = 1}^N\frac{1}{\sigma_2o_t(\bth)}
g\big({z_2(t)},{z_1(t)},x_1(t)x_2(t)\rho\big),
\end{align}
where $z_i(t)=v(t)x_i(t)/\sigma_i$. 
Furthermore, since
\begin{equation}\label{derivative3}
    \frac{\partial\mathcal{L} (\X;\bth)}{\partial \sigma_{12}}
    =\sum_{t = 1}^N \frac{x_1(t)x_2(t)}{\sigma_1\sigma_2o_t(\bth)}f\left(z_1(t),z_2(t)|\rho\right).
\end{equation}
the iterative procedure is
\begin{align}\label{joint_mle_iteration}
\hat{\bth}^{(k+1)}=\hat{\bth}^{(k)} + \mu^{(k)} \left. \frac{\partial \mathcal{L}(\X;\bth)}{\partial \bth}\right|_{\bth=\hat{\bth}^{(k)}},
\end{align}
where $\mu^{(k)}$ is the learning rate at the $k$th iteration.

However, when the number of sub-intervals $l$ is small, i.e., the number of different thresholds is small, the above iterative process can be omitted with minimal performance loss and the estimates are given by those in previous sections. This assertion is proved in Appendix \ref{appendix:joint_MLE} and, in the next section, it is also verified by numerical simulations.


\subsection{Complex-Valued Case}
We now assume $\x$ follows a multivariate complex Gaussian distribution with covariance matrix $\bSi_{\x}$. We perform the widely linear transformation~\cite{Schreier2010}, namely, stacking the real and imaginary parts of $\x$ as $\underline{\x}=[\w^T,\z^T]^T$,
where $\w=\Re(\x)$ and $\z=\Im(\x)$. Then, the covariance matrix of $\underline{\x}$ is
\begin{align}
\bSi_{\underline{\x}}=
\begin{bmatrix}
\bSi_{\w\w}~~  \bSi_{\w\z}\\
\bSi_{\z\w}~~ \bSi_{\z\z}
\end{bmatrix}.
\end{align}
Accordingly, we perform the same procedure to transform the one-bit samples $\y$ into $\underline{\y}$.
Then, $\bSi_{\underline{\x}}$ is estimated from  $\underline{\y}$ via the algorithm in the previous subsection.
Finally, we reconstruct the covariance matrix of $\x$ from $\hat{\bSi}_{\underline{\x}}$ as
\begin{align}
\hat{\bSi}_{\x}=\hat{\bSi}_{\w\w}+\hat{\bSi}_{\z\z}+\imath(\hat{\bSi}_{\z\w}-\hat{\bSi}_{\w\z}).
\end{align}

\subsection{Performance Analysis of the Estimator}

This section analyzes the MSE of the proposed time-varying threshold-based approach. To proceed, we need the following lemma for the asymptotic behavior of the MLE~\cite{Kay_estimation}.

\begin{lemma}\label{lemma_mle}
Under the regularity condition that
\begin{equation}
    \label{eq:regularity}
    \mathbb{E}\left[ \frac{\partial\mathcal{L}(\X;\bth)}{\partial \bth}  \right]=\0,
\end{equation}
the MSE matrix of the MLE can be asymptotically ($N \rightarrow \infty$) approximated by
\begin{align}
    \Q=\F^{-1}(\bth).
\end{align}
where $\F(\bth)$ is the Fisher information matrix (FIM):
\begin{equation}\label{FIM_definition}
\F(\bth)=\mathbb{E}\left[ \frac{\partial\mathcal{L}(\X;\bth)}{\partial \bth}\frac{\partial\mathcal{L}(\X;\bth)}{\partial \bth^T}  \right].
\end{equation}
\end{lemma}
In our case, we have $\bth=[\sigma_1,\sigma_2,\sigma_{12}]^T$. Since the samples are mutually independent, we can compute the Fisher information contributed by each sample separately. Using the first-order derivatives in \eqref{derivative1}-\eqref{derivative3} and the fact that $\x(t)\in\{\pm 1,\pm 1\}$, for $t=1,\cdots, N$, the FIM is computed as
\begin{equation}\label{FIM}
\F(\bth)=\sum_{t=1}^N\sum_{\x(t)\in\{\pm 1,\pm 1\}} o_t(\bth) \left[ \frac{\partial\mathcal{L}(\x(t))}{\partial \bth}\frac{\partial\mathcal{L} (\x(t))}{\partial \bth^T}  \right].
\end{equation}
Appendix \ref{appendix:D} proves that \eqref{eq:regularity} holds. Then, according to Lemma \ref{lemma_mle}, the asymptotic MSE of the elements are obtained as the diagonal entries of $\F^{-1}(\bth)$.

\section{Numerical Results}
\label{sec:simulations}

In this section, we conduct numerical simulations to compare the proposed recovery scheme with existing results using constant~\cite{Liu2021} and random~\cite{Eamaz2022} thresholds. Additionally, we validate the accuracy of our MSE analysis. Each result represents a Monte Carlo simulation based on $10^5$ independent tests.

\subsection{Comparison of Mean Squared Errors}

\begin{figure}[!t]
  \includegraphics[width=0.9\columnwidth]{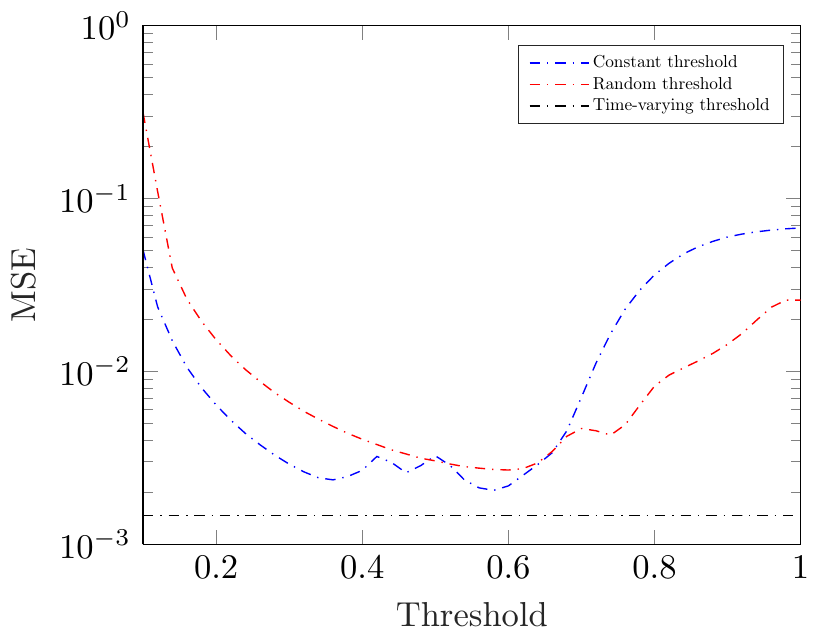}
\caption{Mean squared error versus threshold value
}\label{fig2}
\end{figure}

We commence our comparison by examining the MSE of the proposed recovery technique and the constant and random threshold methods. In Fig. \ref{fig2}, the population parameters are chosen as $\sigma_1=0.25$, $\sigma_2=0.6,$ and $\sigma_{12}=-0.08$, and the number of samples is $N = 1000$. Our approach employs a threshold that varies from $0.1$ to $1$, with increments of $0.1$, and each value is maintained for $1/10$ of the acquisition period. The constant threshold approach takes a different value between 0.1 and 1 for each simulation. For the random threshold method, the thresholds are combined with a dithering signal following $\mathcal{N}(\0_2,0.15\cdot\I_2)$. The results show that the time-varying threshold provides a lower MSE than any constant threshold value, as it can effectively estimate parameters over a wider range. It also outperforms the random threshold approach as it exploits the exact values of the threshold rather than their statistical properties.

\subsection{Influence of Correlation Coefficient}

Next, we examine the impact of the correlation coefficient on estimation accuracy. We set $\sigma_1=0.25$ and $\sigma_2=0.6$, while the correlation coefficient ranges from $-0.95$ to $0.95$, and the number of samples is still $N = 1000$. The constant threshold approach employs a threshold value  of $0.5$,  while the dithering signal corresponding to the random threshold approach and the threshold for our approach remain as in the previous experiment.

\begin{figure}[!t]
  \includegraphics[width=0.9\columnwidth]{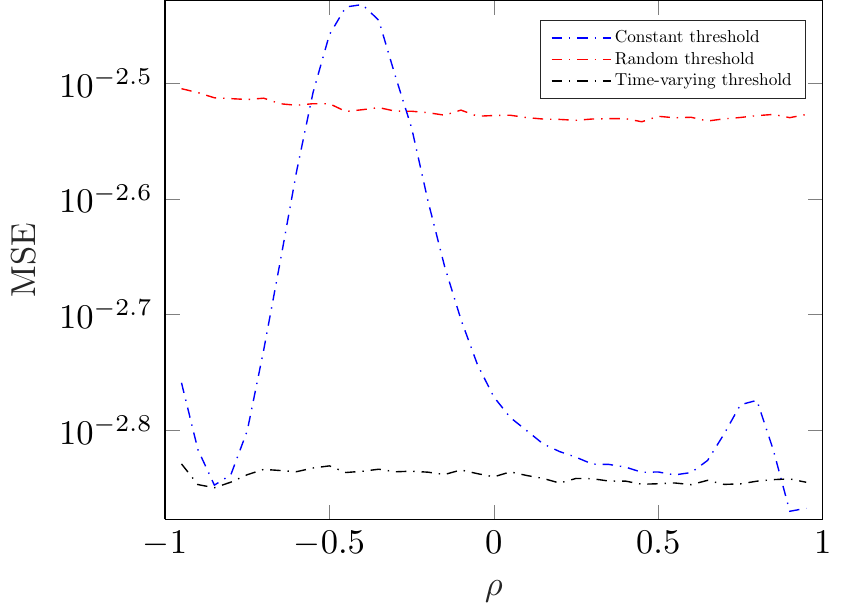}
\caption{Mean squared error versus correlation coefficient}\label{fig3}
\end{figure}

Compared to fixed or random thresholds, our method generally yields smaller MSE and demonstrates greater robustness, as shown in Fig. \ref{fig3}. The dithering approach is also more stable than the constant threshold although it yields a higher MSE on average.

\subsection{Influence of Variance Unevenness }

As illustrated in Fig. \ref{fig1}, the optimal threshold for variance estimation is approximately 1.6 times the standard deviation. Therefore, different variances will make the estimation more challenging for a constant threshold. In the next experiment, we set  $\sigma_1=0.6 + \delta$ and $\sigma_2=0.6 - \delta$. The correlation coefficient is set to $0.5$ and $N = 1000$.

\begin{figure}[!t]
  \includegraphics[width=0.9\columnwidth]{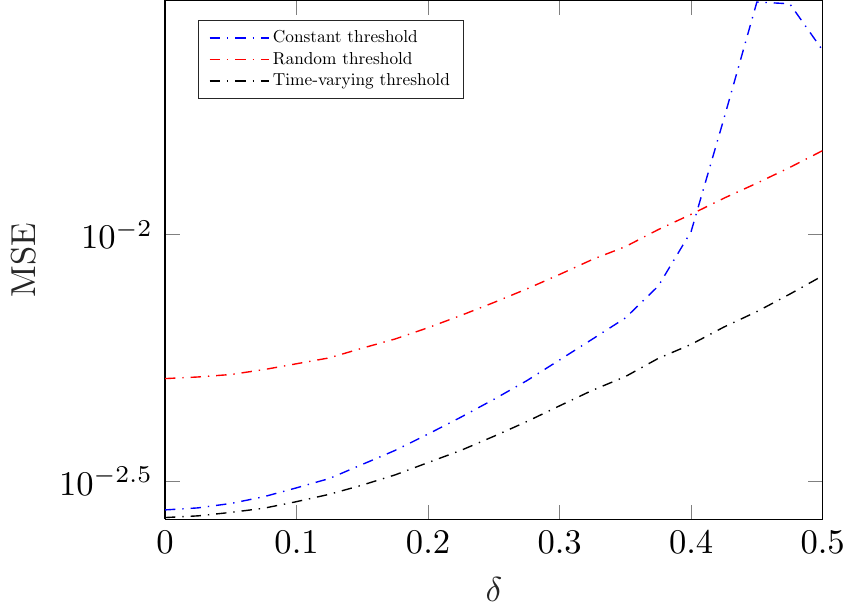}
\caption{Mean squared error versus variance separation level}\label{fig4}
\end{figure}

It is clear that all three approaches experience degradation in performance as the level of unevenness increases, as Fig. \ref{fig4} shows. However, the time-varying threshold approach demonstrates the smallest increase in estimation error, which highlights its robustness when estimating covariance matrices with diverse parameters, which is a common in real-world applications.

\subsection{Influence of the Joint MLE}

In this subsection, we verify the effectiveness of estimating variances separately versus seeking the joint MLE. We collect the largest gradients that emerged in the iteration process in \eqref{joint_mle_iteration} and compare the MSE with and without this process. The results are presented in Table \ref{table1} for $\sigma_1=0.25, \sigma_2=0.6, \rho = 0.5$, and $N = 1000$. We observe that even the largest gradients exhibit negligible values, indicating that the iteration process for joint MLE has a minimal impact on the estimation result. Furthermore, the initial estimates provide nearly identical MSE values as the joint MLE, implying that the iteration process for the joint MLE can be safely omitted without any adverse effects on performance as shown in Appendix \ref{appendix:joint_MLE}.

\begin{table}[!t]
\begin{center}
\caption{Absolute initial gradient and MSE comparison between joint and separate MLEs}\label{table1}
\begin{tabular}{ccc}
   \toprule[1.5pt]
   & \!\!\!\!\!\!\!\!Largest gradient &\!\!\!\!\!\!\!\!\!\!\!\! MSE (Separate)~~~~~ MSE (Joint)  \\\midrule[1pt]
   \!\!$\sigma_1$ & ~~~$8.382\times10^{-3}$ ~~~ &$2.291\times10^{-4}$ ~~~~$2.241\times10^{-4}$~~~~\!\\
   \!\!$\sigma_2$ & ~~~$9.213\times10^{-4}$ ~~~ &$1.024\times10^{-3}$ ~~~~$1.023\times10^{-3}$~~~~\! \\
    \!\!$\sigma_{12}$ & ~~~$8.496\times10^{-7}$ ~~~ &$2.160\times10^{-4}$ ~~~~$2.137\times10^{-4}$~~~~\! \\\bottomrule[1.5pt]
\end{tabular}
\end{center}
\end{table}

\subsection{Theoretical Mean Squared Error}

\begin{figure}[!ht]
\begin{minipage}[b]{0.9\linewidth}
  \centering{\includegraphics{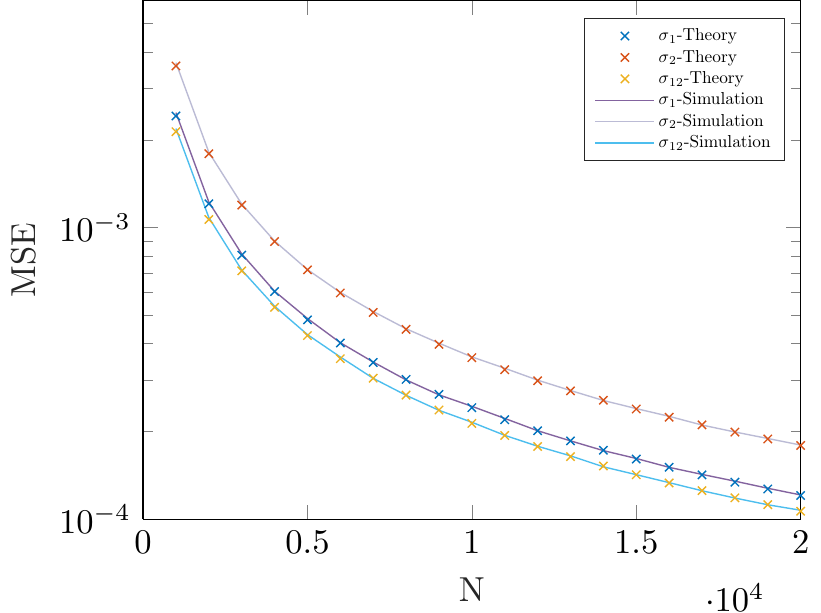}}
\centerline{\small{(a) Time-varying threshold}}
\medskip
\end{minipage}
\hfill
\begin{minipage}[b]{0.9\linewidth}
  \centering{\includegraphics{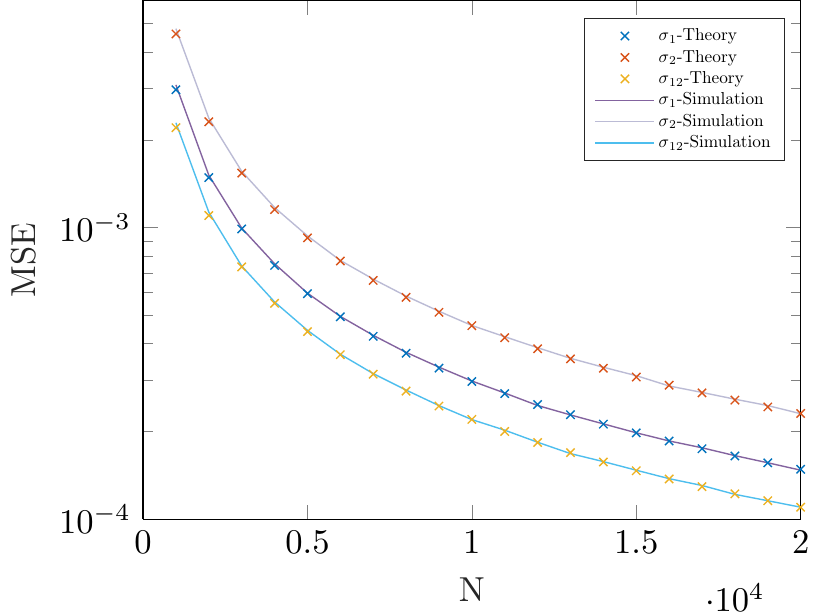}}
\centerline{\small{(b) Constant threshold}}
\medskip
\end{minipage}
\caption{Mean squared error versus number of samples}\label{fig5}
\end{figure}

Now we examine the accuracy of the theoretical MSE of the variance estimator and covariance estimator obtained by inverting the FIM in \eqref{FIM}. The population parameters are set as $\sigma_1=0.8$, $\sigma_2=0.9$, $\sigma_{12}=0.25$, and $N = 1000$. We begin by investigating the theoretical performance of our approach in Fig. \ref{fig5} (a), where the sampling thresholds remain unchanged as previously. The result corresponding to the constant threshold is illustrated in Fig \ref{fig5} (b). It is worth noting that the covariance matrix of the dithering signal in the random threshold approach can be incorporated into that of the signal part, thus, the performance of the random threshold approach is predictable by the result of the constant threshold approach, eliminating the need for a different simulation.

\subsection{DOA Estimation of Coherent sources}

\begin{figure*}[t]
\begin{minipage}[c]{0.3\linewidth}
  \centerline{\includegraphics[width=0.8\linewidth]{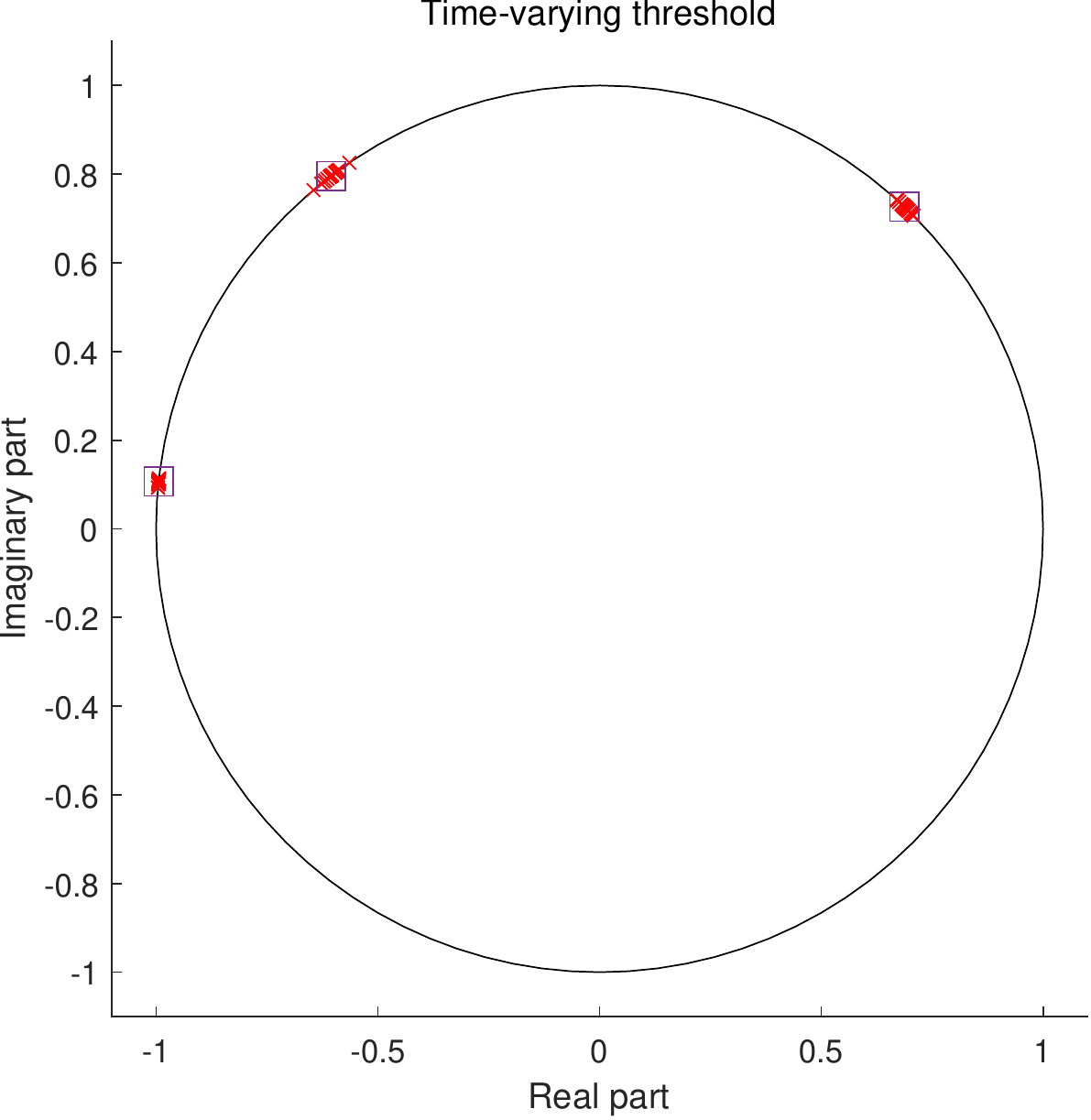}}
 \centerline{\small{a. Time-varying threshold}}
  \end{minipage}
\begin{minipage}[c]{0.3\linewidth}
  \centerline{\includegraphics[width=0.8\linewidth]{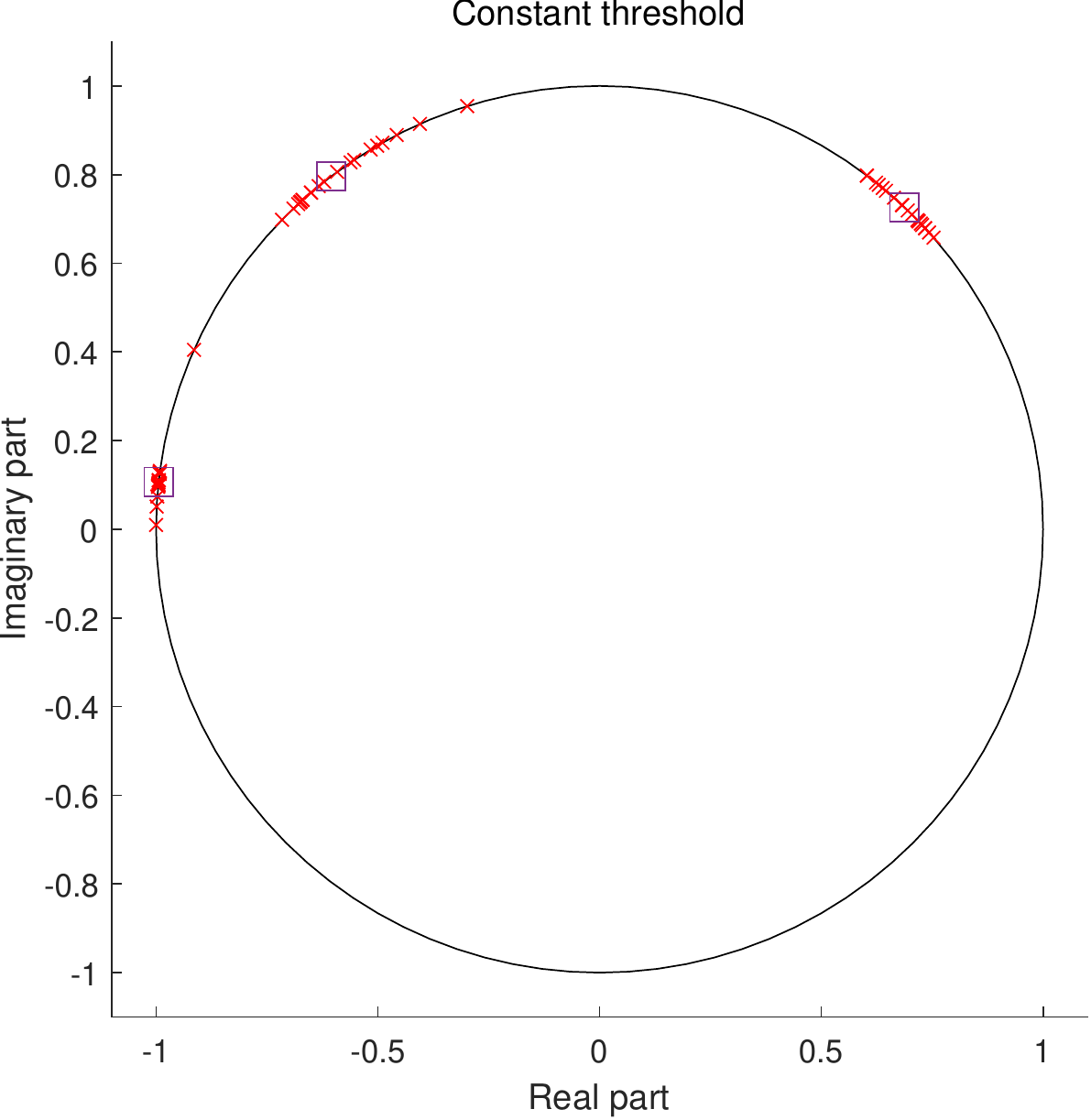}}
\centerline{\small{b. Constant threshold}}
  \end{minipage}
\begin{minipage}[c]{0.3\linewidth}
  \centerline{\includegraphics[width=0.8\linewidth]{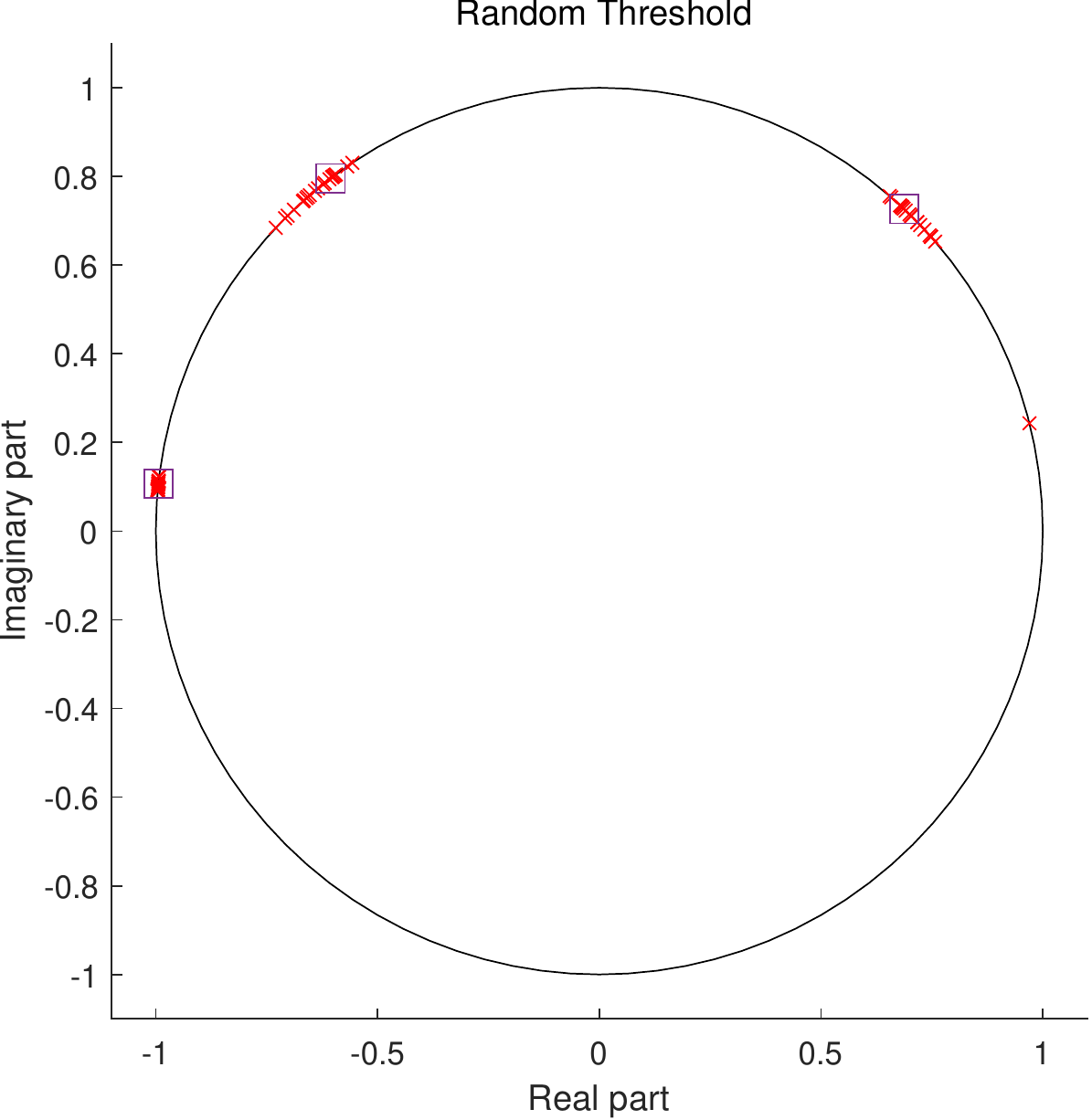}}
\centerline{\small{c. Random threshold}}
  \end{minipage}
\caption{Comparison of estimated DOA} \label{fig6}
\end{figure*}

Finally, we assess the performance of the three methods in a real-world application, specifically the DOA estimation of coherent sources. The covariance matrix is first reconstructed using each of the three methods, and then processed by the EPUMA~\cite{Qian2016} algorithm. A total of $6$ antennas are utilized and there are three sources located at $15^\circ,45^\circ$, and $75^\circ$, with a signal-to-noise-ratio (SNR) of $20$dB. The number of samples is $10000$, and a total of $20$ simulations were conducted. Fig. \ref{fig6} shows that our time-varying threshold approach provides the most accurate and reliable results compared to the constant threshold and random threshold methods. This is due to the fact that the parameters of the actual covariance matrix can span a wide range, making robustness a crucial factor in ensuring estimation precision.

\section{Conclusion}
\label{sec:conclusions}


The results of this paper demonstrate the importance of threshold selection in one-bit estimation of covariance matrices. By examining the limitations of a static threshold approach, a novel time-varying threshold-based recovery scheme is  developed to achieve improved accuracy in the estimation of covariance matrices. The superior performance is demonstrated through both theoretical analysis and numerical simulations, and the results show significantly reduced MSE and enhanced robustness in complex scenarios. This study opens the door for future research to further optimize the threshold selection based on the derived theoretical results of the MSE. The results of this study also have a wide range of potential applications in many areas, including array processing and communications.

\appendices

\section{Proof of \eqref{expansion_var}}
\label{appendix:A}

Computing the first- and second-order derivatives, we have
\begin{align}\label{derivatives}
h'(p_i)&=-\frac{v}{[Q^{-1}(p_i)]^2}\frac{\partial Q^{-1}(p_i)}{\partial p_i},\\
h''(p_i)&=\frac{v}{[Q^{-1}(p_i)]^3}\!\(\!2 \!\left[\frac{\partial Q^{-1}(p_i)}{\partial p_i}\right]^2\!\!\!-\!Q^{-1}(p_i)\frac{\partial^2 Q^{-1}(p_i)}{\partial p_i^2}\!\). \label{derivatives2}
\end{align}
Using the formulas of the derivative of inverse functions, we have:
\begin{align}
\frac{\partial Q^{-1}(a)}{\partial a}&=\frac{1}{Q'(Q^{-1}(a))},\\
\frac{\partial^2 Q^{-1}(a)}{\partial a^2}&=\frac{Q''(Q^{-1}(a))}{[Q'(Q^{-1}(a))]^3}.
\end{align}
Taking into consideration that
\begin{align}\label{q_inv}
Q^{-1}(p_i)=\frac{v}{\sigma_i},
\end{align}
and
\begin{align}
Q'(a) &= \frac{\partial Q(a)}{\partial a} = -\frac{1}{\sqrt{2\pi}}\exp\(-\frac{a^2}{2}\),\\
Q''(a) &= \frac{\partial^2 Q(a)}{\partial a^2} = \frac{a}{\sqrt{2\pi}}\exp\(-\frac{a^2}{2}\),
\end{align}
the derivatives become
\begin{align}\label{inv_de}
\frac{\partial Q^{-1}(p_i)}{\partial p_i}&= -\sqrt{2\pi}\exp\(\frac{v^2}{2\sigma_i^2}\),\\
\frac{\partial^2 Q^{-1}(p_i)}{\partial p_i^2}&= \frac{2\pi v}{\sigma_i}\exp\(\frac{v^2}{\sigma_i^2}\). \label{inv_de_bis}
\end{align}
Substituting \eqref{inv_de}, \eqref{inv_de_bis}, and \eqref{q_inv} into \eqref{derivatives}-\eqref{derivatives2} yields,
\begin{align}
h'(p_i)
&=\frac{\sqrt{2\pi}\sigma_i^2}{v}\exp\(\frac{v^2}{2\sigma_i^2}\),\\
h''(p_i)&=\exp\(\frac{v^2}{\sigma_i^2}\) \left(\frac{4\pi\sigma_i^3}{v^2} - 2\pi \sigma_i \right).
\end{align}

\section{Proof of Lemma \ref{lemma:taylor_expansion}}
\label{appendix:taylor_expansion}
We first establish the first-order Taylor's expansion $p_{12}$ at $\hat{p}_{12}$:
\begin{align}\label{p_ij expansion}
\hat{p}_{12}&=p_{12}(\hat{\sigma}_1,\hat{\sigma}_2,\hat{\sigma}_{12})\Nn\\
&=p_{12}(\sigma_1,\sigma_2,\sigma_{12})
+\frac{\partial p_{12}}{\partial \sigma_1}(\hat\sigma_1-{\sigma}_1)\Nn\\
&\phantom{=}+\frac{\partial p_{12}}{\partial \sigma_2}(\hat\sigma_2-{\sigma}_2)
+\frac{\partial p_{12}}{\partial \sigma_{12}}(\hat\sigma_{12}-{\sigma}_{12}).
\end{align}
Rearranging terms, we have:
\begin{multline}\label{p_ij expansion2}
\sigma_{12}-\hat{\sigma}_{12}
= \\ \frac{\partial \sigma_{12}}{\partial p_{12}}\[p_{12}-\hat{p}_{12}-\frac{\partial p_{12}}{\partial \sigma_1}(\sigma_1-\hat{\sigma}_1)-\frac{\partial p_{12}}{\partial \sigma_2}(\sigma_2-\hat{\sigma}_2)\],
\end{multline}
where we have used the inverse function rule. In the previous subsection, we obtained
\begin{align}\label{sigma_i expansion}
\sigma_i-\hat{\sigma}_i=h'(p_i)(p_i-\hat{p}_i)+\mathcal{O}((p_i-\hat{p}_i)^2),~~i=1,2.
\end{align}
Combining \eqref{p_ij expansion2} and \eqref{sigma_i expansion} we have the following linear function:
\begin{align}
\sigma_{12}-\hat{\sigma}_{12}=&\frac{\partial \sigma_{12}}{\partial p_{12}}\[p_{12}-\hat{p}_{12}-\frac{\partial p_{12}}{\partial \sigma_1}h'(p_1)(p_1-\hat{p}_1)\right.\Nn\\
&\left.-\frac{\partial p_{12}}{\partial \sigma_2}h'(p_2)(p_2-\hat{p}_2)\]\Nn\\
=&\l\[p_1-\hat{p}_1,p_2-\hat{p}_2,p_{12}-\hat{p}_{12}\]^T.
\end{align}
Moreover, since
\begin{align}
p_{12}=\int_{\frac{v}{\sigma_1}}^\infty\int_{\frac{v}{\sigma_2}}^\infty f\left(x_1,x_2|\rho\right)dx_1dx_2,
\end{align}
the partial derivative $\partial p_{12}/\partial\sigma_1$ is computed via the following integration:
\begin{align}
\frac{\partial p_{12}}{\partial \sigma_1}
&=\frac{v}{\sigma_1^2} \int_{\frac{v}{\sigma_2}}^\infty f\left(\frac{v}{\sigma_1},x_2\Big|\rho\right)dx_2\Nn\\
&\phantom{=}-\frac{\rho}{\sigma_1}\frac{\partial }{\partial \rho}\int_{\frac{v}{\sigma_1}}^\infty\int_{\frac{v}{\sigma_2}}^\infty f\left(x_1,x_2|\rho\right)dx_1dx_2\Nn\\
&=\frac{v}{\sigma_1^2} \frac{1}{\sqrt{2\pi}} \exp\(-\frac{v^2}{2\sigma_1^2}\)Q\left(\frac{v/\sigma_2-\rho v/\sigma_1}{\sqrt{1-\rho^2}}\right)\Nn\\
&\phantom{=}-\frac{\rho}{\sigma_1}f\left(\frac{v}{\sigma_1},\frac{v}{\sigma_2}\Big|\rho\right)\Nn\\
&=\frac{1}{\sigma_1}g\(\frac{v}{\sigma_1},\frac{v}{\sigma_2},\rho\)
\end{align}
where we have used Leibniz integral rule and also \eqref{derivative_rho} to compute the last term. Similarly, we could obtain $\partial p_{12} / \partial \sigma_2$.
Finally, and using again the derivative $\partial p_{12}/\partial\rho$ that has been calculated in \eqref{derivative_rho}, it is straightforward to obtain
\begin{align}
\frac{\partial p_{12}}{\partial \sigma_{12}}=\frac{\partial p_{12}}{\partial \rho}\frac{\partial \rho}{\partial \sigma_{12}}
=&\frac{1}{\sigma_1\sigma_2}f\(\frac{v}{{\sigma}_1},\frac{v}{{\sigma}_2}\Big|\rho\).
\end{align}

\section{Proof of \eqref{R}
\label{appendix:C}}

As $\hat{p}_1$, $\hat{p}_2$ and $\hat{p}_{12}$ are scaled binomial random variables, the diagonal entries of $\R$ are easily determined as:
\begin{align}
[\R]_{1,1}&=\frac{p_1-p_1^2}{N},\\
[\R]_{2,2}&=\frac{p_2-p_2^2}{N},\\
[\R]_{3,3}&=\frac{p_{12}-p_{12}^2}{N}.
\end{align}

The covariance between $\hat{p}_1$ and $\hat{p}_{12}$ is
\begin{align}
\mathbb{C}(\hat{p}_1,\hat{p}_{12})&=\mathbb{E}[\hat{p}_1\hat{p}_{12}]-p_1p_{12}\Nn\\
&=\mathbb{E}[N_1N_{12}]/N^2-p_1p_{12}
\end{align}
where $N_1=N\hat{p}_1$ and $N_{12}=N\hat{p}_{12}$.
The value of $\mathbb{E}[N_1N_{12}]$ is
\begin{align}
\mathbb{E}[N_1N_{12}] &= \sum_{k,l=1}^N\mathbb{E}\left[\frac{x_1(k) + 1}{2} \frac{x_1(l) + 1}{2} \frac{x_2(l) + 1}{2}\right]\Nn\\
&=\sum_{k=1}^N\mathbb{E}\left[\left(\frac{x_1(k) + 1}{2}\right)^2 \frac{x_2(k) + 1}{2}\right] \Nn \\
&\phantom{=}+\sum_{\substack{k,l=1\\k\neq l}}^N\mathbb{E}\left[\frac{x_1(k) + 1}{2} \frac{x_1(l) + 1}{2} \frac{x_2(l) + 1}{2}\right]\Nn\\
&=Np_{12}+N(N-1)p_1p_{12} \Nn\\
&=N^2 p_1 p_{12} + Np_{12}(1 - p_1),
\end{align}
where we have used the independence between $x_1(k)$ and $x_2(l)$. Therefore, we have
\begin{align}
\mathbb{C}(\hat{p}_1,\hat{p}_{12})=\frac{p_{12}(1-p_{1})}{N}.
\end{align}
Similarly, we can obtain
\begin{align}
\mathbb{C}(\hat{p}_2,\hat{p}_{12})=\frac{p_{12}(1-p_{2})}{N}.
\end{align}
Finally, since $\mathbb{E}[N_1N_{2}]$ is
\begin{align}
\mathbb{E}[N_1N_{2}] &= \sum_{k,l=1}^N\mathbb{E}\left[\frac{x_1(k) + 1}{2} \frac{x_2(l) + 1}{2}\right]\Nn\\
&=\sum_{k=1}^N\mathbb{E}\left[\frac{x_1(k) + 1}{2} \frac{x_2(k) + 1}{2}\right] \Nn \\
&\phantom{=}+\sum_{\substack{k,l=1\\k\neq l}}^N\mathbb{E}\left[\frac{x_1(k) + 1}{2} \frac{x_2(l) + 1}{2}\right]\Nn\\
&=Np_{12}+N(N-1)p_1p_2,
\end{align}
where we have used again the independence between $x_1(k)$ and $x_2(l)$, the last covariance is
\begin{align}
\mathbb{C}(\hat{p}_1,\hat{p}_{2})&=\mathbb{E}[N_1N_{2}]/N^2-p_1p_{2}\Nn\\
&=\frac{p_{12}-p_1p_2}{N}.
\end{align}
The proof is complete.

\section{Proof of the vanishing gradient with small number of sub-intervals}
\label{appendix:joint_MLE}

Let us denote the original estimates by $\hat{\bth}$, obtained in Sections \ref{sec:rec_diagonal} and \ref{sec:rec_nondiagonal}, and the joint MLE by $\hat{\bth}'$, obtained in Section \ref{sec:joint_rec} after the gradient-based algorithm converges. We start by considering the first sub-interval, which is of length $n$ and define the following random variables:
\begin{align}
K_{1}&=\frac{\sum_{t=1}^n [x_1(t)+1][x_2(t)+1]}{4n},\\
K_{2}&=\frac{\sum_{t=1}^n [x_1(t)+1][x_2(t)-1]}{4n},\\
K_{3}&=\frac{\sum_{t=1}^n [x_1(t)-1][x_2(t)-1]}{4n},\\
K_{4}&=\frac{\sum_{t=1}^n [x_1(t)-1][x_2(t)+1]}{4n},
\end{align}
which estimate the probability of $\x(t) = \boldsymbol{\epsilon}_i$, with
\begin{align}
\begin{array}{c}
\boldsymbol{\epsilon}_1=[+1,+1]^T,~~~\boldsymbol{\epsilon}_2=[+1,-1]^T,\\
\boldsymbol{\epsilon}_3=[-1,-1]^T,~~~\boldsymbol{\epsilon}_4=[-1,+1]^T.
\end{array}
\end{align}
Then, the derivative of the log-likelihood with respect to $\sigma_1$ evaluated at the original estimate is
\begin{align}
     \left.\sum_{t=1}^n\frac{\partial\mathcal{L}(\x(t);\bth)}{\partial \sigma_1}\right|_{\bth = \hat{\bth}}       &= \left.\sum_{i=1}^4 nK_i\frac{\partial\mathcal{L}(\x=\boldsymbol{\epsilon}_i;\bth)}{\partial \sigma_1} \right|_{\bth = \hat{\bth}}\Nn\\
    &=\sum_{i=1}^4\frac{nK_i}{q_i}g\(\frac{v(t)\boldsymbol{\epsilon}_{i,1}}{\hat{\sigma}_1},\frac{v(t)\boldsymbol{\epsilon}_{i,2}}{\hat{\sigma}_2},\hat{\rho}\),
 \end{align}
 where $\hat{\rho} = \hat{\sigma}_{12}/\hat{\sigma}_1 \hat{\sigma}_2$, and
\begin{align}
q_i=\int_{\frac{\tau_iv_1(t)}{\hat{\sigma}_1}}^\infty\int_{\frac{\tau_iv_2(t)}{\hat{\sigma}_2}}^\infty f\left(y_1,y_2\Big|\tau_i\hat{\rho}\right)dy_1dy_2,
\end{align}
is the probability that $\x=\boldsymbol{\epsilon}_i$, with $\tau_i=\mathbf{\epsilon}_{i,1}\mathbf{\epsilon}_{i,2}$.
     Recalling the definition of function $g(z_1,z_2,\rho)$ in \eqref{g_def}, it is easily seen that
\begin{equation}\label{g_function}
g(\kappa_1,\kappa_2,\varrho)=-g(-\kappa_1,\kappa_2,-\varrho).
\end{equation}
Therefore, we have
\begin{multline}\label{likelihood_minus}
       \left.\sum_{t=1}^n\frac{\partial\mathcal{L}(\x(t);\bth)}{\partial \sigma_1}\right|_{\bth = \hat{\bth}} =n\(\frac{K_1}{q_1}-\frac{K_4}{q_4}\)g\(\frac{v(t)}{\hat{\sigma}_1},\frac{v(t)}{\hat{\sigma}_2},\hat{\rho}\)\\
      + n\(\frac{K_2}{q_2}-\frac{K_3}{q_3}\)g\(\frac{v(t)}{\hat{\sigma}_1},-\frac{v(t)}{\hat{\sigma}_2},
      -\hat{\rho}\).
\end{multline}

Since $(nK_1,nK_2,nK_3,nK_4)$ follows a multinomial distribution with probabilities $(q_1,q_2,q_3,q_4)$, the random variables $\frac{K_1}{q_1},\frac{K_2}{q_2},\frac{K_3}{q_3},\frac{K_4}{q_4}$ follow asymptotically a Gaussian distribution $\mathcal{N}(\1_4,\C)$,
where
\begin{align}
    [\C]_{i,j}=
\begin{cases}
\frac{1-q_i}{nq_i}, &i=j\\
-\frac{1}{n}, &i\neq j.
\end{cases}
\end{align}
Then, we have
\begin{align}
\frac{K_1}{q_1}-\frac{K_4}{q_4}&=\mathcal{O}\(n^{-\frac{1}{2}}\),\\
\frac{K_2}{q_2}-\frac{K_3}{q_3}&=\mathcal{O}\(n^{-\frac{1}{2}}\),
\end{align}
 and \eqref{likelihood_minus} becomes
\begin{align}
\left.\sum_{t=1}^n\frac{\partial\mathcal{L}(\x(t);\bth)}{\partial \sigma_1}\right|_{\bth = \hat{\bth}} = \mathcal{O}\(n^{\frac{1}{2}}\).
\end{align}
Note, that this derivative is not zero because $\hat{\sigma}_1$ was obtained using the likelihood of $x_1(t), t = 1, \ldots, N$.
To proceed, we apply a first-order Taylor's expansion
to the derivative of the log-likelihood, which results in
\begin{multline}
    \left.\sum_{t=1}^N\frac{\partial\mathcal{L}(\x(t);\bth)}{\partial \sigma_1}\right|_{\bth = \hat{\bth}'} = \left.\sum_{t=1}^N\frac{\partial\mathcal{L}(\x(t);\bth)}{\partial \sigma_1}\right|_{\bth = \hat{\bth}}\\
    +(\hat{\sigma}_1'-\hat{\sigma}_1)\left.\sum_{t=1}^N\frac{\partial^2\mathcal{L}(\x(t);\bth)}{\partial \sigma_1^2}\right|_{\bth = \hat{\bth}}.
\end{multline}
Since $\hat{\sigma}_1'$ is the solution to the equation
\begin{align}
\left.\sum_{t=1}^N\frac{\partial\mathcal{L}(\x(t);\bth)}{\partial \sigma_1}\right|_{\bth = \hat{\bth}'} = 0,
\end{align}
we have
\begin{align}\label{update_term}
    \hat{\sigma}_1'-\hat{\sigma}_1&\approx - \frac{\displaystyle \left.\sum_{t=1}^N\frac{\partial\mathcal{L}(\x(t);\bth)}{\partial \sigma_1}\right|_{\bth = \hat{\bth}}}{\displaystyle \left.\sum_{t=1}^N\frac{\partial^2\mathcal{L}(\x(t);\bth)}{\partial \sigma_1^2}\right|_{\bth = \hat{\bth}}}.
\end{align}
Now we investigate the second-order derivative. When $n$ is large, we have
\begin{equation}
\left.\sum_{t=1}^n\frac{\partial^2\mathcal{L}(\x(t);\bth)}{\partial \sigma_1^2}\right|_{\bth = \hat{\bth}}
\rightarrow \\ n \mathbb{E}\[\left.\frac{\partial^2\mathcal{L}(\x(t);\bth)}{\partial \sigma_1^2}\right|_{\bth = \hat{\bth}}\], 
\end{equation}
which is of order $n$ since $\mathbb{E}\[\partial^2\mathcal{L}(\x(t);\bth)/\partial \sigma_1^2|_{\bth = \hat{\bth}}\] = \mathcal{O}\(1\)$.
Therefore, the numerator in \eqref{update_term} is a summation of $l$ terms of order $n^{\frac{1}{2}}$ while the denominator is a summation of $l$ terms of order $n$, where $l$ is the number of sub-intervals. As a result, we obtain
\begin{equation}
    \hat{\sigma}_1'-\hat{\sigma}_1 \approx \mathcal{O}\(n^{-\frac{1}{2}}\).
\end{equation}
This implies that when $l$ is small and $n=N/l$ is large, the estimated $\sigma_1$ in the joint MLE is close to the initial estimate. Similarly, we can obtain $\hat{\sigma}_2'-\hat{\sigma}_2 \approx \mathcal{O}(n^{-\frac{1}{2}})$. Furthermore, since $\hat{\sigma}_{12}$ is obtained using the two-channel data by solving
\begin{equation}
    \left.\frac{\partial{\mathcal{L}(\X;\hat{\sigma}_1,\hat{\sigma}_2,{\sigma}_{12})}}{\partial{\sigma_{12}}}\right|_{\sigma_{12} = \hat{\sigma}_{12}}=0,
\end{equation}
its initial gradient is already $0$. With $\hat{\sigma}_1$ and $\hat{\sigma}_2$ remaining almost unchanged, the gradient of ${\sigma}_{12}$ is also negligible. Then, the original estimate $\hat{\bth}$ and the joint MLE by $\hat{\bth}'$ are close.


\section{Proof of \eqref{eq:regularity}}
\label{appendix:D}

We first prove that, for each sample vector $\x(t)$, $(t=1,\cdots, N)$,  the regularity condition holds, namely:
\begin{equation}
    \mathbb{E}\left[ \frac{\partial\mathcal{L}(\x(t);\bth)}{\partial \bth}  \right]=\0.
\end{equation}
Then the result naturally holds for the collection of all samples. At first, we have
\begin{align}\label{sigma_1_sum}
    &\mathbb{E}\left[ \frac{\partial\mathcal{L}(\x(t);\bth)}{\partial \sigma_1}  \right]\Nn\\
    &=\sum_{\x(t)\in\{\pm 1,\pm 1\}} o_t(\bth)\frac{\partial\mathcal{L}(\x(t);\bth)}{\partial \sigma_1}\Nn\\
    &=\sum_{\x(t)\in\{\pm 1,\pm 1\}} \frac{1}{\sigma_1} g\(\frac{v(t)x_1(t)}{\sigma_1},\frac{v(t)x_2(t)}{\sigma_2},x_1(t)x_2(t)\rho\).
\end{align}
Taking into account \eqref{g_function}, we have
\begin{align}
    \left.\frac{\partial\mathcal{L}(\x(t);\bth)}{\partial \sigma_1}\right|_{\x(t)=\boldsymbol{\epsilon}_1}= \left.- \frac{\partial\mathcal{L}(\x(t);\bth)}{\partial \sigma_1}\right|_{\x(t)=\boldsymbol{\epsilon}_4},\Nn\\
        \left. \frac{\partial\mathcal{L}(\x(t);\bth)}{\partial \sigma_1}\right|_{\x(t)=\boldsymbol{\epsilon}_2}=\left. - \frac{\partial\mathcal{L}(\x(t);\bth)}{\partial \sigma_1}\right|_{\x(t)=\boldsymbol{\epsilon}_3},
\end{align}
which yields $\mathbb{E}\left[\partial\mathcal{L}(\x(t);\bth)/\partial \sigma_1 \right] = 0$. This verifies the regularity condition for $\sigma_1$, which can be easily extended to $\sigma_2$.
Similarly, we have
\begin{align}\label{sigma_12_sum}
    \mathbb{E}&\left[ \frac{\partial\mathcal{L}(\x(t);\bth)}{\partial \sigma_{12}}  \right]\Nn\\
    &=\sum_{\x(t)\in\{\pm 1,\pm 1\}} o_t(\bth) \frac{\partial\mathcal{L}(\x(t);\bth)}{\partial \sigma_{12}}\Nn\\
    &=\sum_{\x(t)\in\{\pm 1,\pm 1\}}\frac{x_1(t)x_2(t)}{\sigma_1\sigma_2}f\left(\frac{v(t)x_1(t)}{\sigma_1},\frac{v(t)x_2(t)}{\sigma_2}\Big|\rho\right).
\end{align}
Since
\begin{align}
  f\left(z_1,z_2|\rho\right)=f\left(-z_1,z_2|-\rho\right),
\end{align}
following the same process as above, we can prove that the summation in \eqref{sigma_12_sum} is $0$. This concludes the proof.

\bibliographystyle{IEEEtran}


\end{sloppypar}


\begin{thebibliography}{10}

\bibitem{Balevi2019}
E. Balevi and J. G. Andrews,
``One-bit OFDM receivers via deep learning,''
\emph{ IEEE Trans. Commun.},
vol. 67, no. 6, pp. 4326–4336, Jun. 2019.

\bibitem{Zhang2020}
Y. Zhang, M. Alrabeiah and A. Alkhateeb,
``Deep learning for massive MIMO with 1-Bit ADCs: when more antennas need fewer pilots,''
\emph{IEEE Wireless Commun. Lett.},
vol. 9, no. 8, pp. 1273–1277, Aug. 2020.

\bibitem{Mo2015}
J. Mo and R. W. Heath,
``Capacity analysis of one-bit quantized MIMO systems with transmitter channel state information,''
\emph{IEEE Trans. Signal Process.},
vol. 63, no. 20, pp. 5498–5512, Oct. 2015.

\bibitem{Qian2019SPL}
C. Qian, X. Fu, and N.~D.~Sidiropoulos,
``Amplitude retrieval for channel estimation of MIMO systems with one-bit ADCs,''
\emph{IEEE Signal Process. Lett.},
vol. 26, no. 11, pp. 1698-1702, Nov. 2019.

\bibitem{Li2017}
Y. Li, C. Tao, G. Seco-Granados, A. Mezghani, A. L. Swindlehurst,
and L. Liu,
``Channel estimation and performance analysis of one-bit massive MIMO systems,''
\emph{IEEE Trans. Signal Process.},
vol. 65, no. 15, pp. 4075–4089, Aug 2017.

\bibitem{Choi2016TC}
J. Choi, J. Mo, and R. W. Heath,
``Near maximum-likelihood detector and channel estimator for uplink multiuser massive MIMO systems with one-bit ADCs,''
\emph{IEEE Trans. Commun.},
vol. 64, no. 5, pp. 2005–2018, May 2016.

\bibitem{Shalom2002TAES}
O. Bar-Shalom and A. J. Weiss,
``DOA estimation using one-bit quantized measurements,''
\emph{IEEE Trans. Aerosp. Electron. Syst.},
vol. 38, no. 3, pp. 868-884, Jul. 2002.

\bibitem{Yu2016SPL}
K. Yu, Y. D. Zhang, M. Bao, Y. Hu, and Z. Wang,
``DOA estimation from one-bit compressed array data via joint sparse representation,''
\emph{IEEE Signal Process. Lett.},
vol. 23, no. 8, pp. 1279-1283, Sep. 2016.

\bibitem{Liu2017ICASSP}
C. L. Liu and P. P. Vaidyanathan,
``One-bit sparse array DOA estimation,''
in \emph{Proc. IEEE Int. Conf. Acoust., Speech, Signal Process.},
New Orleans, LA, USA, Mar. 2017, pp. 3126-3130

\bibitem{Stein2016ITGWSA}
M. Stein, K. Barbe, and J. A. Nossek,
``DOA parameter estimation with 1-bit quantization bounds, methods and the exponential replacement,''
in \emph{Proc. 20th Int. ITG Workshop Smart Antennas},
Munich, Germany, 2016, pp. 1-6.

\bibitem{Sedighi2020}
S. Sedighi, B. Shankar, M. Soltanalian, and B. Ottersten, ``One-bit DoA
estimation via sparse linear arrays,''
in \emph{Proc. IEEE Int. Conf. Acoust. Speech Signal Process.},
Barcelona, Spain, 2020, pp. 9135–9139.

\bibitem{Sedighi2021}
S. Sedighi, B. S. Mysore R, M. Soltanalian and B. Ottersten,
``On the performance of one-bit DoA estimation via sparse linear arrays,''
\emph{IEEE Trans. Signal Process.},
vol. 69, pp. 6165-6182, 2021.

\bibitem{Ren2017}
J. Ren and J. Li,
``One-bit digital radar,''
in \emph{Proc. IEEE Asil. Conf. on Sig., Sys., and Comp.}, Pacific Grove, USA, 2017, pp. 1142–1146.

\bibitem{Xiao2022}
Y. -H. Xiao, D. Ramírez, P. J. Schreier, C. Qian and L. Huang,
``One-bit target detection in collocated MIMO radar and performance degradation analysis,''
\emph{IEEE Trans. Veh. Technol.},
vol. 71, no. 9, pp. 9363-9374, Sept. 2022

\bibitem{Xi2020TAES}
F. Xi, Y. Xiang, Z. Zhang, S. Chen, and A. Nehorai,
``Joint angle and Doppler frequency estimation for MIMO radar with one-bit sampling:
A maximum likelihood-based method,''
\emph{IEEE Trans. Aerosp. Electron. Syst.}, 2020.

\bibitem{Xi2020TSP}
F. Xi, Y. Xiang, S. Chen, and A. Nehorai,
``Gridless parameter estimation for one-bit MIMO radar with time-varying thresholds,''
\emph{IEEE Trans. Signal Process.},
vol.~68, pp. 1048-1063, 2020.

\bibitem{Liu2021IET}
B.~Liu, B.~Chen, M.~Yang,
``Parameter estimation and CRB analysis of 1-bit colocated MIMO radar,''
\emph{IET Radar Sonar Navigat.,},
vol.~1, no.~13, pp. 1-13, Mar. 2021.


\bibitem{Jin2020TAES}
B. Jin, J. Zhu, Q. Wu, Y. Zhang, and Z. Xu,
``One-bit LFMCW radar: Spectrum analysis and target detection,''
\emph{IEEE Trans. Aerosp. Electron. Syst.},
vol.~56, no.~4, pp. 2732-2750, Aug. 2020.

\bibitem{Stein2015TSP}
M. Stein, A. Kurzl, A. Mezghani, and J. A. Nossek,
``Asymptotic parameter tracking performance with measurement data of 1-bit resolution,''
\emph{IEEE Trans. Signal Process.},
vol. 63, no. 22, pp. 6086-6095, Nov. 2015.

\bibitem{Qian2016}
C. Qian, L. Huang, N. D. Sidiropoulos and H. C. So,
``Enhanced PUMA for direction-of-arrival estimation and its performance analysis,'' \emph{IEEE Trans. Signal Process.},
vol.~64, no.~16, pp. 4127-4137, Aug. 2016.

\bibitem{Wei2012a}
\textup{L. Wei and O. Tirkkonen},
``Spectrum sensing in the presence of multiple primary users,''
\emph{IEEE Trans. Commun.}, vol. 60, no. 5, pp. 1268-1277, May 2012.

\bibitem{Xiao2018a}
Y. Xiao, L. Huang, J. Xie, and H. C. So,
``Approximate asymptotic distribution of locally most powerful invariant test for independence: Complex case,''
\emph{IEEE Trans. Inf. Theory}, vol.~64, no.~3, pp. 1784-1799, Mar.~2018.

\bibitem{Zhao2021a}
Y. Zhao, X. Ke, B. Zhao, Y. Xiao and L. Huang,
``One-Bit Spectrum Sensing Based on Statistical Covariances: Eigenvalue Moment Ratio Approach,''
\emph{IEEE Wireless Communications Letters}, vol.~10, no.~11, pp. 2474-2478, Nov. 2021.

\bibitem{Xiao2018b}
 Y. Xiao, L. Huang, J. Zhang, J. Xie, and H. C. So,
``Performance analysis of locally most powerful invariant test for sphericity of Gaussian vectors in coherent MIMO radar,''
\emph{IEEE Trans. Veh. Technol.}, vol. 67, no. 7, pp. 5868-5882, Jul. 2018.

\bibitem{Liu2015TAES}
W. Liu, Y. Wang, J. Liu, W. Xie, H. Chen, and W. Gu,
``Adaptive detection without training data in colocated MIMO radar,''
\emph{IEEE Trans. Aerosp. Electron. Syst.},
vol. 51, no. 3, pp. 2469-2479, Jul. 2015.



\bibitem{Vleck1966}
J. H. Van Vleck and D. Middleton,
``The spectrum of clipped noise,''
\emph{Proc. IEEE},
vol. 54, no. 1, pp. 2-19, Jan. 1966.

\bibitem{Bussgang1952}
J. J. Bussgang, ``Cross-correlation function of amplitude-distorted
Gaussian signals,'' Tech. Rep. 216, Res. Lab. Elec.,
Mas. Inst. Technol., March 1952.

\bibitem{Minkoff}
J. Minkoff, ``The role of AM-to-PM conversion in
memoryless nonlinear systems,''
\emph{IEEE Trans. Commun.},
vol. 33, no. 2, pp. 139–144, 1985.

\bibitem{Liu2021}
C.-L. Liu and Z.-M. Lin, “One-bit autocorrelation estimation with nonzero thresholds,”
in \emph{Proc. IEEE Int. Conf. Acoust. Speech Signal Process.}, Toronto, Canada,
2021, pp. 4520–4524.

\bibitem{Price1958}
R. Price, ``A useful theorem for nonlinear devices having Gaussian
inputs,''
\emph{IRE Trans. Inf. Theory},
vol. 4, no. 2, pp. 69–72, June 1958.

\bibitem{Jacovitti1994}
G. Jacovitti and A. Neri, ``Estimation of the autocorrelation function of complex Gaussian stationary processes by amplitude clipped signals,''
\emph{IEEE Trans. Inf. Theory},
vol. 40, no. 1, pp. 239–245, Jan. 1994.

\bibitem{Eamaz2022}
A. Eamaz, F. Yeganegi and M. Soltanalian,
``Covariance recovery for one-bit sampled non-stationary signals with time-varying sampling thresholds,''
\emph{IEEE Trans. Signal Process.},
vol. 70, pp. 5222-5236, 2022.

\bibitem{Zhao2019}
B. Zhao, L. Huang and W. Bao,
``One-bit SAR imaging based on single-frequency thresholds,''
\emph{IEEE Trans. Geosci. Remote Sens.},
vol. 57, no. 9, pp. 7017-7032, Sept. 2019

\bibitem{Knudson2016}
K. Knudson, R. Saab and R. Ward, ``One-Bit compressive sensing With norm estimation,''
\emph{IEEE Trans. Inf. Theory},
vol. 62, no. 5, pp. 2748-2758, May 2016.

\bibitem{Liu2020}
C.-L. Liu and P. P. Vaidyanathan,
``One-bit normalized scatter matrix estimation for complex elliptically symmetric distributions,''  in \emph{Proc. IEEEInt. Conf. Acoust. Speech Signal Process.},
Barcelona, Spain, 2020, pp. 9130-9134.

\bibitem{Kay_estimation}
S. M. Kay, \emph{Fundamentals of Statistical Signal Processing: Estimation Theory}.
NJ: Prentice-Hall, 1993.

\bibitem{Knoblauch2008}
A. Knoblauch,
``Closed-form expressions for the moments of the Binomial probability distribution,'' SIAM Journal on Applied Mathematics, vol. 69, no. 1, pp. 197-204,

\bibitem{Schreier2010}
P. Schreier and L. Scharf, \emph{Statistical Signal Processing of Complex-Valued Data: The Theory of Improper and Non-Circular Signals}. Cambridge, U.K.: Cambridge Univ. Press, 2010.







\end{thebibliography}

\end{document}